\documentclass[a4paper, 10pt, 3p, number, sort&compress, times]{elsarticle}

\usepackage{graphicx}
\usepackage{dcolumn}
\usepackage{bm}
\usepackage{latexsym}
\usepackage{amsfonts}
\usepackage{amssymb}
\usepackage{amsmath}
\usepackage{bbold}
\usepackage{ulem}
\usepackage[utf8]{inputenc} 
\usepackage{xspace}
\usepackage{booktabs}
\usepackage[colorlinks=true, urlcolor=blue, linkcolor=red]{hyperref}
\usepackage{xcolor}

\definecolor{darkspringgreen}{rgb}{0.09, 0.45, 0.27}

\usepackage{xspace}
\newcommand{\ISDM}{ICM\xspace}
\newcommand{\Dref}{D_{\rm ref}}

\usepackage{algorithm}
\usepackage{algpseudocode}

\begin{document}
\date{\today}
\title{A fast and accurate method for inferring solid-state diffusivity in lithium-ion battery active materials: improving upon the classical GITT approach}
\author[1]{A. Emir G\"umr\"uk\c{c}\"uo\u{g}lu \corref{cor1}}
\ead{emir.gumrukcuoglu@port.ac.uk}
\author[1]{James Burridge}
\author[2]{Kieran O'Regan}
\author[3,4]{Emma Kendrick}
\author[1,4]{Jamie M. Foster}

\cortext[cor1]{Corresponding author}

\affiliation[1]{organization={School of Mathematics and Physics,
University of Portsmouth},
addressline={Lion Terrace},
city={Portsmouth},
 citysep={}, 
postcode={PO1 3HF},
country={UK}}

\affiliation[2]{organization={About:Energy},
addressline= {Labs Camden - Atrium The Stables Market, Chalk Farm Road},
city={London},
citysep={},
postcode={NW1 8A},
country={UK}}

\affiliation[3]{organization={School of Metalurgy and Materials,
University of Birmingham},
addressline= {Edgbaston},
city={Birmingham},
citysep={},
postcode={B15 2TT},
country={UK}}

\affiliation[4]{organization={The Faraday Institution},
addressline= {Quad One, Becquerel Avenue, Harwell Campus},
city={Didcot},
citysep={},
postcode={OX11 0RA},
country={UK}}

\begin{abstract}

Data collected using the galvanostatic intermittent titration technique (GITT) and application of the Sand equation is a ubiquitous method for inferring the solid-state diffusivity in lithium-ion battery {active materials}. However, the experiment is notoriously time-consuming and the Sand equation relies on assumptions whose applicability can be questionable.
We propose a novel methodology, termed Inference from a Consistent Model (\ISDM),
which enables inference of solid-state diffusivity using the same physical model employed for prediction, and is applicable to more general and quick-to-measure data.
We infer the diffusivity (as a function of inserted lithium concentration) by minimising the residual sum of squares between data and solutions to a spherically-symmetric nonlinear diffusion model in a single representative active material particle. Using data harvested from the NMC cathode of a commercial LG M50 cell we demonstrate that the \ISDM is robust,
and yields more accurate diffusivity estimates, while relying on data that are five times faster to collect than that required by the classical approach. Moreover, there is good reason to believe that further speed ups could be achieved when other types of data are available.
This work contributes towards developing faster and more reliable techniques in parameter inference for lithium-ion batteries, and the code required to deploy \ISDM is provided to facilitate its adoption in future research.

\end{abstract}
\maketitle

\section{Introduction}
\label{sec:intro}
The Doyle-Fuller-Newman (DFN) model \cite{doyle93,fuller94,fuller94b,1994FuDoNe} is widely acknowledged as the gold standard in physics-based modelling for lithium-ion batteries and offers a framework for understanding and predicting device behaviour at the level of individual and electrode pairs. Provided that the parameterisation has been performed accurately, the DFN model has been shown to reliably predict the electrochemical response of real batteries in realistic operating conditions; with observed and predicted voltages often matching to within a single percentage point \cite{zulke2021parametrisation,schmitt2023full,ecker2015parameterization,ecker2015parameterizationb}. So ubiquitous is the DFN model that a wealth of literature exists on various model simplifications and extensions, many of which are discussed in a recent review by Brosa Planella et. al \cite{BrosaPlanella2022}. One prevalent class of simplified DFN models are the so-called single particle models (SPM) \cite{moura2016battery,guo2010single,marquis2019asymptotic,richardson2020generalised}.
These are generally accurate for low to moderate C-rates  (below 1C, or thereabouts), where all the particles in each electrode behave similarly, allowing the model to be reduced to lithium transport within a single ``representative'' particle per electrode.
Regardless of whether a DFN model is being used, or one of its simplified counterparts, the heart of most physics-based battery models is a spherically-symmetric diffusion equation which is taken to describe the distribution of inserted lithium within the particles of electrode active material. The predictive power of the DFN, and related physics-based models, is predicated upon knowledge of parameters many of which cannot be directly measured but must instead be inferred through physical principles and indirect experimental observations. Therefore, the process of parameterisation -- the accurate inference of these parameters -- is crucial for developing effective battery models, which in turn are key to advancing device improvement and control \cite{Wang_2022}.

One of the most influential parameters in physics-based models is the solid-state diffusivity of lithium within the active materials, $D(c)$, which varies (often rather strongly) with the concentration of inserted lithium, $c$. This diffusivity cannot be measured directly and is therefore indirectly inferred using one of several methods. {In all classical inference approaches, it is assumed that there is only one diffusing species (the inserted lithium) and that the diffusion coefficient is constant over a certain time period. Measurements are often repeated at different concentrations to map out how the diffusivity varies with concentration. Transport is taken to follow Ficks law of diffusion; any kinetics related to phase transformations are not accounted for, and typically, volume changes are ignored. In electrochemical systems the measurable voltage is a function of the surface concentration at the interface with the electrolyte, and therefore the properties at the surface of the material are being probed rather than those of its bulk.
Thus, any inferred diffusion coefficient should be regarded as an \textit{effective} property rather than a true solid-state diffusivity; for brevity, we shall refer to it simply as the diffusivity.
Provided that the inference technique is consistent with the model subsequently used for prediction, it is this effective diffusivity that becomes the quantity of practical interest, i.e.~in developing robust and predictive models.

Established techniques include chronoamperometry, in which a voltage is applied and the current response is measured. The potential is stepped from a value at which the species of interest does not undergo redox reaction to one at which the current becomes diffusion-controlled \cite{em1}.
The potentiostatic intermittent titration technique (PITT) is an extension of this, in which a sequence of voltage steps are applied. The Cottrell equation (see e.g. \cite{Bard_2001_book}) can be used to estimate the effective diffusion coefficient from each voltage transient. After each current transient, a rest period is introduced to observe the open-circuit voltage, from which the ion concentration can be estimated.
When a current is applied, an overpotential is typically observed due to Ohmic and charge-transfer resistance; this overpotential is minimised if a low current is used.
Electrochemical voltage spectroscopy (EVS) is a faster approach to PITT \cite{bark}, where the current is allowed to decay to a low value, the rest step is removed, and any overpotential observed is assumed negligible.
The most widely used method is based upon chronopotentiometry, which involves applying a constant current and observing a voltage transient, from which the diffusion coefficient can be calculated using the Sand equation.
In the galvanostatic intermittent titration technique (GITT), a rest period is introduced after each current pulse to allow the voltage to relax to open circuit, and the effective diffusion coefficient can then be extracted \cite{Weppner1977}.
Each rest step typically lasts 1–3 hours to allow the system to reach equilibrium, and as a result, these tests can often take several weeks to complete.
More recently, an intermittent current interruption technique (ICI) has been proposed \cite{PMID:37085556,yin}; here, a constant current is applied with a pulsed interruption period. During each rest, a voltage transient is observed, and the Sand equation can be applied to extract a diffusion coefficient. In ICI, the overpotential can be subtracted from the measured voltage to estimate the open circuit voltage (OCV). Although the diffusion coefficient reported from ICI often agrees with that from GITT, the time constants used are much shorter than is typical. It is likely that solid-state diffusion is not the only process being captured at these shorter timescales and rest periods (typically 5 seconds).
Other techniques include cyclic voltammetry (CV), where a voltage sweep is applied and the current response is measured. This technique is only applicable to reversible reactions and does not readily yield diffusion coefficients as a function of concentration. Very low sweep rates are typically required for solid-state battery electrodes due to their slow kinetics \cite{kim2020}.
Electrochemical impedance spectroscopy (EIS) can be used to estimate the diffusion coefficient from the low-frequency response, typically via the Warburg element \cite{yu1999}.
Some of these techniques, such as EIS and CV, rely on separating different physical processes by their characteristic time constants. However, this assumption can lead to errors in parameter estimation, particularly when time scales overlap or the system response is not well resolved \cite{lee2022}.
To improve accuracy in electrochemical tests, overpotentials are minimised by using thin electrodes, where mass transfer limitations can be neglected.
Particles are assumed to be spherical, with uniform porosity and 100\% electrochemically active surface area.
In contrast, real-world data often come from thicker electrodes, where electrolyte transport in the pores also impacts observed kinetics and diffusion \cite{chen2022}, and where particle size distributions are often non-uniform or bimodal.

Of the above, perhaps the most widespread inference approach is to collect data with the GITT, where the electrode is driven with a small constant current for a short time, then allowed to relax until the inserted lithium is fully diffused (at uniform concentration), with the forcing--relaxation cycle repeated until the concentration (or equivalently state-of-charge or stoichiometry) range of interest is covered. To a reasonable approximation, this allows each forcing--relaxation cycle to be analysed separately as an independent diffusion problem with constant diffusivity and homogeneous initial concentration. Using the analytic solution of a semi-infinite slab (or, the \textit{Sand equation} \cite{doi:10.1080/14786440109462590}), one can determine a diffusion constant corresponding to the small range of concentrations in each pulse--relax period \cite{Weppner1977}.

Despite its ubiquity, the GITT and Sand equation approach presents several limitations. A key limitation lies in the questionable assumptions embedded in the Sand equation. Inference is carried out using a solution derived for semi-infinite slabs, yet the
predictive models (such as the DFN and its simplifications) are based on diffusion equations in a sphere. Thus there is inconsistency between the inference model and the model used for prediction. The semi-infinite slab approximation is appropriate for pulse durations $t_{\rm pulse}$ much shorter than the diffusion time scale $t_d = R^2/D$. The accuracy decreases as the ratio $t_{\rm pulse}/t_d$ increases, with a 5\% deviation from the spherical solution at $t_{\rm pulse}/t_d \approx3\times 10^{-2}$ \cite{Nickol_2020}. For $t_{\rm pulse}>0.04\,t_d$, the Sand equation becomes inadequate for determining $D(c)$ (a detailed discussion is presented in \ref{app:SlabtoSphere}).
This result imposes a practical limitation on the experiment design. The duration of the constant pulse should be no longer than $4\%$ of the diffusion time, an upper limit that can significantly extend total experiment duration. As a consequence, in order to adequately reconstruct $D(c)$ across the relevant range of concentrations, it is necessary to apply low currents several hundred times. Each pulse is followed by a relaxation period, typically lasting 1--3 hours, to restore uniform concentration, a prerequisite for the validity of the Sand equation in the next pulse. Thus the total data collection time can span weeks.

To enable a later comparison with the assumptions embedded in \ISDM, we now make explicit the assumptions embedded in the Sand equation. These are
\begin{itemize}
	\item[i] A single particle can be taken as a representative of all particles within the electrode; i.e. all particles behave almost identically.
	\item[ii] The transport process within the particle is governed by spherically-symmetric diffusion.
	\item[iii] Reaction overpotentials are negligible compared to the open-circuit voltage.
	\item[iv] The diffusion lengthscales during a single pulse-relax period are significantly smaller than the particle radii, such that the curvature of the spherical geometry becomes negligible. The surface can therefore be locally approximated as a planar slab.
\end{itemize}

In this paper, we propose a novel approach to infer the diffusivity as a function of inserted lithium concentration, which we term \textit{inference from a consistent model} (\ISDM). The key idea is to use an inference method that is directly motivated by the models in which the inferred parameters will subsequently be used (e.g., the DFN model and its simplified counterparts). Specifically, we solve the fully nonlinear spherical diffusion model in a single representative particle for a known surface lithium flux (directly proportional to the known current applied to/drawn from the cell), and determine the concentration at the particle surface. Via the equilibrium overpotential, $U_{\rm eq}(c)$, we calculate the cell voltage $V(t)$, and by comparing this quantity to the measured cell voltage, we infer $D(c)$.
In contrast with the classical approach, the \ISDM offers the following benefits:
(i) it can be applied to data with \textit{any} current profile, provided the concentration spans the range of values of interest and the dominant contribution to the cell voltage is the open circuit voltage, thereby allowing for very significantly reduced data collection times;
(ii) the inference model is fully compatible with the models used for prediction, eliminating internal inconsistencies and ensuring validity across all diffusion timescales;
(iii) as we will show, the parameters inferred with \ISDM are objectively more accurate than those inferred by classical means.

The remainder of this paper is organised as follows. In \S\ref{sec:model}, we outline the mathematics underlying the \ISDM method.
In \S\ref{expdat} we summarise the LG M50 experimental dataset used in our study and describe the derivation of reference quantities.
We demonstrate self-consistency of the \ISDM, and lack of self-consistency for the classical approach, in \S\ref{sec:validation}. In \S\ref{sec:validation_GITT} we compare the diffusivities inferred by applying both the classical method and our \ISDM method to classical GITT pulse-relax data; with comparable performance being demonstrated. In \S\ref{sec:validation_constant} we showcase the generality of our approach by accurately inferring diffusivity entirely from galvanostatic data. Then, in \S\ref{sec:comparison}, we compare the classical and \ISDM  approaches in recovering a known diffusivity using synthetic data obtained by solving a fully-fledged DFN model. The superior performance of the \ISDM method here demonstrates its objective advantages over the classical method. We conclude with \S\ref{sec:conclusions} where we summarise and discuss our results.

\section{Inference from a consistent model: choosing the model}
\label{sec:model}

In this section, we introduce the general procedure adopted in the \ISDM framework and describe the specific model used in our case study. The defining principle of \ISDM is consistency: the same physical model is used both for parameter inference and for prediction.
Our primary aim is to write down the simplest physics-based model that contains only the parameters that one infers with GITT, thereby allowing a comparison of the methods in a truly like-for-like setting.
However, \ISDM itself is not tied to this model, or any particular model. It bears emphasising that if one were confronted with data which could be better explained by a more complicated model (likely involving more parameters) then the general philosophy of the \ISDM methodology remains robust. The same cannot be said for the classical GITT method because it is inherently tied to an analytical approximation of a particular model via the Sand equation.

\subsection{A single particle model (SPM)}
\label{sec:spm}
We consider a representative particle with radius $R$, where the molar concentration of inserted lithium, $c(t,r)$, in the sphere follows the radially symmetric diffusion equation
\begin{equation} \label{eq:diffusioneq}
\frac{\partial c}{\partial t} = \frac{1}{r^2}\,\frac{\partial}{\partial r}\left(r^2\,D(c)\,\frac{\partial c}{\partial r}\right)\,,
\end{equation}
where $D(c)$ is the concentration-dependent diffusivity. This partial differential equation (PDE) is subject to the boundary and initial conditions
\begin{equation}
\left.\frac{\partial c}{\partial r}\right\vert_{r=0} = 0\,, \qquad
\left.-D(c)\,\frac{\partial c}{\partial r}\right\vert_{r=R} = \frac{j}{F}\,, \qquad
c|_{t=0} = c_0\,,
\label{eq:midprob}
\end{equation}
where $j(t)$ is the surface current density and $c_0(r)$ is the initial distribution of inserted Li. Under the usual single particle model assumption that the dynamics of each particle are very similar, we can (to a good approximation) apportion the current applied to/drawn from the cell, $I(t)$, equally between particles, such that
\begin{equation} \label{eq:endprob}
j = \pm \frac{I}{n\,4\,\pi\,R^2}\,,
\end{equation}
where $n$ is the number of particles in the electrode, and $F=e\,N_A=9.6485\times10^4 \mathrm{C/mol}$ is Faraday's constant. We adopt the upper sign for an anode and the lower one for a cathode, so that positive values of $I(t)$ correspond to a lithiating cathode (and delithiating anode) and hence battery discharging. In subsequent sections we shall focus on NMC material and will therefore primarily be interested in the lower sign.

In the experiments pertinent to this study, one applies a time-dependent current, $I(t)$, and measures the voltage response. Neglecting reaction overpotentials and internal resistance (arising from the conduction of electrons in the solid matrix and ions in the electrolyte), the only remaining contribution to the voltage is that from the open-circuit voltage so that
\begin{equation} \label{modend}
V_{\text{SPM}} = U_{\rm eq}(c_{\rm surf})\,,
\end{equation}
where $c_{\rm surf}(t) = c(t,r)|_{r=R}$ is the surface concentration of inserted Li. 

Before moving on to describe the inference approach we make explicit the assumptions embedded in (\ref{eq:diffusioneq})-(\ref{modend}):
\begin{itemize}
	\item[1.] A single particle can be taken as a representative of all particles within the electrode.
	\item[2.] The transport process within the particle is spherically symmetric diffusion.
	\item[3.] Reaction overpotentials are negligible compared to the open-circuit voltage.
	\item[4.] Potential drops due to internal resistances are negligible.
\end{itemize}	
Assumptions 1, 2 and 3 coincide exactly with assumptions i, ii and iii in the classical GITT method, see \S\ref{sec:intro}. Assumption 4, appears to be a further restriction to the applicability of the SPM, however it is in fact not a separate assumption but rather is a consequence of 1. A requirement for all particles to behave similarly (assumptions 1 and i) is that they each have similar local environments, i.e. they are immersed in an electrolyte of similar concentration and potential and are also connected to a solid matrix at similar potentials. That is to say that both the electrolyte and electrode must be almost equipotential, i.e. relatively unpolarized. If the liquid and solid phases are almost equipotential then their resistances are negligible in the regime of interest, i.e. 4 pertains. Thus, having seen that 4 follows 1, we can conclude that the assumptions embedded in the SPM are fewer and less restrictive than those that pertain to the classical approach. In particular, the SPM requires no assumptions on the relative sizes of timescales relevant to current excitations vs. those for diffusion, and it is exactly this property that allows the \ISDM with SPM to be applied to \textit{any} set of data that does not violate 1--3.

\subsection{Inference of diffusivity and the parameterisation scheme}

Let $\mathcal{D} = \{t_i, I_i, V_i\}_{i=1}^N$  be a set of experimental data and let us assume that the particle number $n$, radius $R$ and $U_{\rm eq}(c)$ are known. In order to infer a concentration-dependent diffusivity, we must first assume some parametrisation which we denote $D(c ; \theta)$, where $\theta$ is the parameter vector. Representations with few parameters are generally incapable of capturing the complexity of estimates of $D(c)$ extant in the literature
(see e.g. \cite{horner2021, zulke2021parametrisation, IVANISHCHEV2017479}). On the other hand, increasing the number of parameters leads to a high-dimensional decision variable space which makes it impractical to scan for a global minimum of the loss function.
In this study, we represent $D(c)$ as a piecewise linear function, where the entries in the $\mathcal{N}$-dimensional parameter vector $\theta$ represent diffusivity values at knots whose positions are determined from the data, as described in \ref{app:optapp}. Other parameterisations are, of course, possible; the choice here is guided by simplicity and flexibility.

In order to test the plausibility of different potential diffusivities (and therefore different $\theta$) we require solutions to Eqs.~(\ref{eq:diffusioneq})-(\ref{modend}). Owing to the lack of analytical solutions we employ a numerical approach in which the spatial derivatives in \eqref{eq:diffusioneq} are treated using the conservative control-volume method \cite{Zeng2013} thereby reducing the PDE and its boundary and initial conditions to a system of coupled ordinary differential equations (ODE) for the concentrations at given collocation points, $r_i$, as functions of time only
\footnote{For real data (Sections \ref{sec:validation_GITT} and \ref{sec:validation_constant}), we use $101$ radial grid points. In Section \ref{sec:comparison}, where DFN synthetic data was generated with $30$ radial points, the \ISDM approach adopts the same discretisation for consistency}.
This system of coupled ODEs can then be integrated in time using one of many ODE integrators; throughout this study we used LSODA wrapper for \texttt{solve\_ivp} with \texttt{rtol=1e-5} and \texttt{atol=1e-8}. These settings aim to control the local error in each step to approximately five significant digits, though the actual global accuracy may vary, particularly in stiff regimes.
As we shall see later this is sufficient accuracy to ensure that errors engendered by the discretisation and numerical time-stepping do not materially affect the value of the loss function. Having carried out the time integration, we obtain the predicted voltage time-series $V^{\text{SPM}}_i=V^{\text{SPM}}(t_i)$ which we can compare to the measured voltage $V_i$ in the data $\mathcal D$. To make this comparison more precise we define the loss function as the mean of the residual sum of squares:
\begin{equation}
{\mathcal{L}(\theta) = \frac{1}{N}\sum_{i=1}^N \left(V_i - V^{\text{SPM}}_i \right)^2}.
\label{eq:totalloss}
\end{equation}
Finding the diffusivity that is most plausible according to the data then amounts to finding the value of $\theta$ that minimises the loss. That is
\begin{equation} \label{optprob}
\hat{\theta} = \underset{\theta}{\rm argmin}\,\mathcal{L}(\theta)\,,
\end{equation}
which is equivalent to a maximum likelihood estimate under the assumption of Gaussian errors. As we will see
below
we typically use $\mathcal{N}=50$ knots in order to capture the features of realistic functions $D(c)$. Thus, the optimisation problem (\ref{optprob}) requires us to search a 50-dimensional parameter space. This is a non-trivial task, but one that can be alleviated using the strategies that we discuss in detail in \ref{app:optapp}.

\subsection{Performance metrics}
To quantify the accuracy of the predictions from the inference approaches we require some performance metrics. In the subsequent sections we will infer diffusivities in two different settings. First, for validation purposes, inference will be carried out for synthetic data that has been generated by forward solving (\ref{eq:diffusioneq})-(\ref{modend}) with a known, ``reference'', diffusivity.
Second, we consider genuine experimental data, where the correct diffusivity cannot be determined directly and can only be inferred.
In the first case, because the solution of the inference problem is known in advance, we can objectively measure the success of the inference as follows:
\begin{equation} \label{eq:R2_D}
{R^2_D(D) = 1 - \frac{\Big\langle \left(D(c) -\Dref(c) \right)^2 \Big\rangle_c}{\Big\langle\left( \Dref(c) - \langle \Dref(c)\rangle_c\right)^2 \Big\rangle_c	}\,,}
\end{equation}
{where $\langle f(c) \rangle_c$ denotes the mean of a function $f(c)$ taken over the domain of concentration values $c$ for which the function is defined.}
\footnote{It is important to note that this measure loses significance if $\Dref$ is flat, i.e. does not exhibit variations with concentration. Nevertheless, for the cases examined in this paper, the diffusivities typically have strong dependence on concentration and $R^2_D$ proves to be a reliable metric for assessing the differences of the inferred diffusivities from the true one.} The metric $R^2_D$ quantifies the proximity of the inferred diffusivity to its reference counterpart, i.e. that which generated the data.

In the second case, when genuine experimental data are used, and the underlying effective diffusivity is unknown, we measure the performance of inference by examining how well the observed voltage can be explained by the inferred diffusivity.
It is well-known that for devices operating at low to moderate C-rates, a significant portion of the voltage response is determined by the open-circuit voltage, $U_{\rm eq}(c)$. Thus, throughout this study we expect model predictions to lie relatively close to the observed data. This motivates measuring model performance not by how much of the voltage is explained by a model with inferred parameters, but rather how much of the voltage {\it beyond that predicted by a null model} (involving only the open circuit voltage) can be captured by a parametrised model. We define our null model to correspond to a solution in which $D(c)\to \infty$, i.e. that with instantaneous diffusion, yielding
\begin{equation}
	V^{\rm null}(t) = U_{\rm eq}(c_{\rm av}(t))\,,
\end{equation}
where $c_{\rm av}(t)$ is the average concentration across the particle at time $t$ and can be computed as described in \eqref{eq:cav_def}. Then, the time series of the excess voltage beyond that of the null model in the experimental data and model prediction are
\begin{equation}
	\Delta V_i = V_i - V^{\rm null}(t_i)\,, \qquad {\Delta V^{\rm SPM}_i=  V^{\rm SPM}(t_i) - V^{\rm null}(t_i)\,,}
\end{equation}
respectively.
Thus, the metric $R_V^2$, measuring the match between the predicted and measured voltage in excess of that explained by the null model is
\begin{equation}
	R_V^2 = 1 - \frac{\sum_{i=1}^N \left(\Delta V_i - \Delta V^{\rm SPM}_i\right)^2}{\sum_{i=1}^N\left(\Delta V_i - \frac1N\,\sum_{j=1}^N\Delta V_j\right)^2}\,.
	\label{eq:R2_definition}
\end{equation}
Note that $R_V^2 \in(-\infty, 1]$, where $R_V^2 = 1$ indicates a perfect explanation of the data, values close to $0$ suggest that the model does not improve upon the null model, and negative values reveal that the model deviates farther from the data than the null model.

To clarify whether the model is evaluated on the same dataset it was inferred from or on a different one, we use the subscripts \textit{train} and \textit{test}.
The former $R^2_{V, ~\mathrm{train}}$ refers to the case where the model is both inferred from and evaluated on the same dataset, providing a measure of goodness of fit. The latter, $R^2_{V, ~\mathrm{test}}$, applies when the model is evaluated on a separate dataset not seen during inference, and quantifies generalisability. The mathematical definition remains the same in both cases.

\section{\label{expdat}Experimental data}
Throughout this study we use experimental data collected from the NMC811 cathode of a commercial LG M50 cell (high energy, $\approx$5mAh/cm$^2$), representative of commercial lithium‑ion devices.
A highly detailed investigation of this device has been reported in  Ref.~\cite{Chen_2020}. What follows is an abridged summary of similar expirements collected from the same device, which we use for the inferences presented in this article.
The data contain four cycles: the first one is a charging cycle with constant current for about 10 hours, followed by a discharge cycle of the same duration. Then there is a {GITT} pulse-relax charge cycle consisting of 249 periods of constant pulse (150 seconds) and relaxation (1 hour). This is similar to, but slightly differs from, that used in \cite{Chen_2020}. The last part of the data is a {GITT} discharge cycle with the same pulse and relaxation durations for 250 periods.
A  coin cell of $15 {\rm ~mm}$ diameter and $75.6 {\rm ~\mu m}$ thickness was used, and the constant current for all cycles is $I=0.78 \mathrm{~mA}$ (corresponding to approximately $C/10$).
The average particle radius is $R=5.22{\rm ~\mu m}$. The delithiation starts at minimum state of charge with fully diffused concentration $c = 0.9084\,c_{\rm max}$ with the maximum concentration determined to be $c_{\rm max} = 51765{\rm ~mol/m^3}$.

Before moving on, we establish a reference diffusivity, $\Dref(c)$ and an open-circuit voltage, $U_{\rm eq}(c_{\textrm{surf}})$, derived using classical methods. The former is inferred by applying the Sand equation to the GITT pulse-relaxation data, while the latter is taken as the voltage measured at the end of the relaxation periods, see Fig.~\ref{fig:self_validation_GITT}. Both quantities are then linearly interpolated across the relevant concentration range for use in solvers.

\section{Self-consistency}
\label{sec:validation}
We start by demonstrating the efficacy of the \ISDM method in accurately reproducing a known diffusivity when using synthetic data. The synthetic data is generated by solving the model (\ref{eq:diffusioneq})-(\ref{modend}) using
the reference diffusivity and open-circuit voltage, with an input current identical to that used in the experimental {GITT} charging data, with 249 periods of pulse and relaxation.

The result of this procedure is shown in Fig.\ref{fig:self_validation_GITT} for $\mathcal{N}=50$ knots. Alongside the diffusivity inferred by the \ISDM, we show both $\Dref$, the reference diffusivity used to generate the synthetic data, and ${D}_{\rm GITT}(c)$ which is the diffusivity obtained by applying the Sand equation to the synthetic data. The corresponding performance metrics are
\begin{eqnarray}
{\rm for \, \, the \, \, \ISDM:} \qquad R^2_{V~\mathrm{train}} = 0.997\,, \qquad 	R^2_D = 0.991\,;\\
{\rm for \, \, classical \, \, GITT:} \qquad R^2_{V~\mathrm{train}} = 0.800\,, \qquad 	R^2_D = 0.882\,.
\end{eqnarray}
The diffusivity estimated by the \ISDM is highly accurate, demonstrating the method's ability to recover a known diffusivity provided that the data has arisen from a model consistent with the inference model itself. This is precisely what one would anticipate: the premise of the \ISDM is to minimise deviations in the voltage as predicted by (\ref{eq:diffusioneq})-(\ref{modend}), and those in the data. Since, in this case, the data was produced by solving (\ref{eq:diffusioneq})-(\ref{modend}) there is every reason to expect an excellent match and, in fact, a lack of one would be an indication of improper implementation. The small deviation of both performance metrics for the \ISDM from unity (a perfect inference) is explained by our choice to represent the diffusivity with $\mathcal{N}=50$ knots; increasing the number of knots allows the performance metrics to increase towards 1. However, this is not surprising and may arguably lead to diminishing returns, given the jagged nature of $D_{\rm ref}$ and the limited resolution of the numerical solver.

The deviation of the performance metrics from unity for the classical approach highlights the inconsistency between the Sand equation and (\ref{eq:diffusioneq})-(\ref{modend}). Since the synthetic data was generated from (\ref{eq:diffusioneq})-(\ref{modend}), the mismatch can be attributed to assumption iv, that the sphere can be approximated as a semi-infinite slab. This is alarming because the classical inference model (the Sand equation) is therefore not consistent with models that are later used to predict battery behaviour. This lack of consistency has been identified previously by other authors, see e.g. \cite{Nickol_2020}, and we show in \ref{app:SlabtoSphere} that the duration of a GITT pulse plays an important role in the accuracy of the slab approximation of the Sand equation, and for the Chen et al data the error exceeds 5\% across most stoichiometries and 7.5\% across half of them.
This result makes lucid the importance of consistency between a model used for inference and that which will be subsequently used to make predictions.

\begin{figure}[ht]
\centering
\includegraphics[width=0.49\columnwidth]{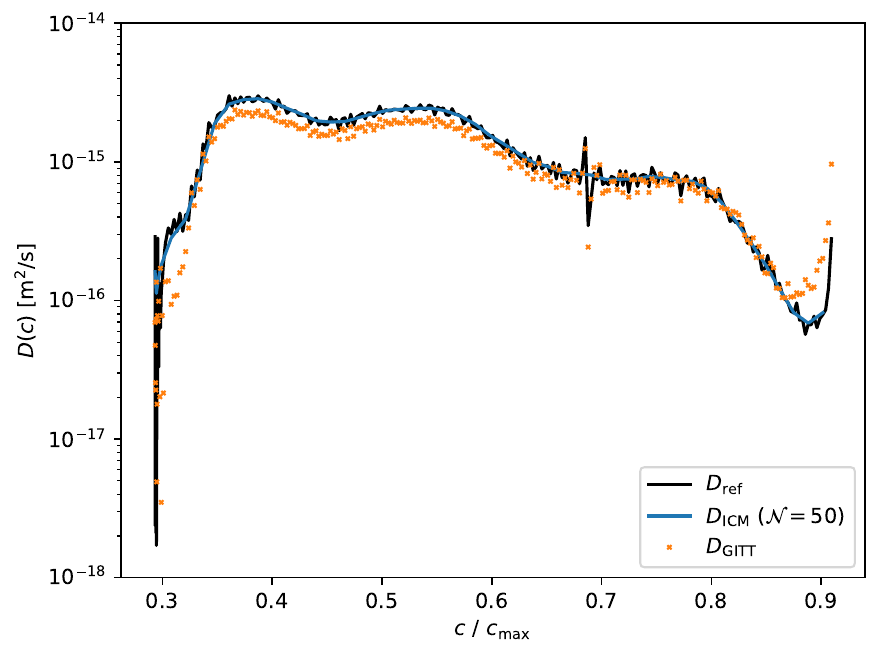}~
\includegraphics[width=0.49\columnwidth]{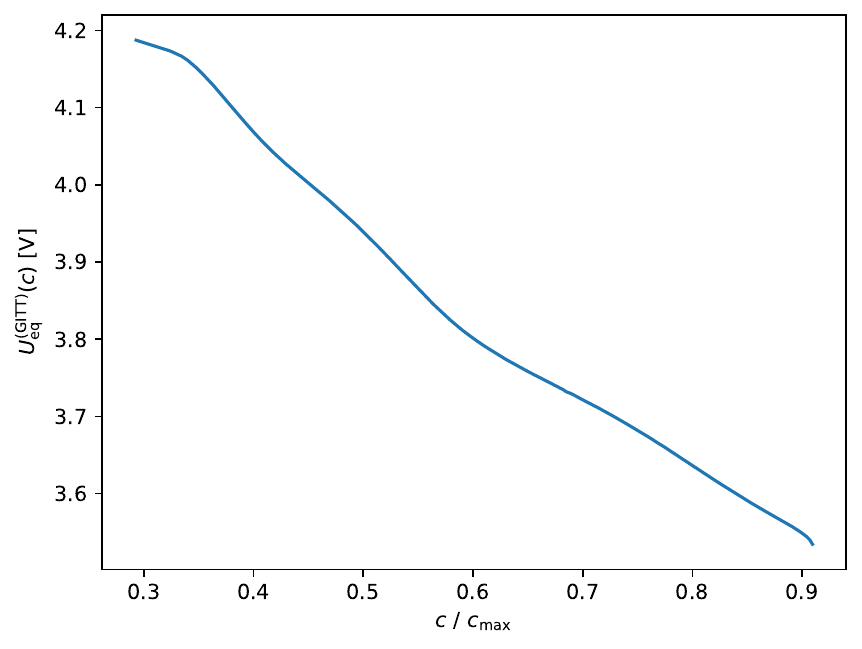}
\caption{
{\bf Left:} Concentration-dependent diffusivity used in and inferred from the synthetic data. The applied current profile matches the GITT charge cycle from \cite{Chen_2020}. The solid black curve shows the reference diffusivity $D_{\text{ref}}(c)$ used to generate the synthetic dataset. The solid blue curve shows the diffusivity inferred from the synthetic data using \ISDM, while the orange markers represent the diffusivity obtained by applying the classical GITT analysis to the same synthetic data.
{\bf Right:} Equilibrium potential $U_{\rm eq}(c)$ used to generate the synthetic data.
In both panels, the reference diffusivity and equilibrium potential were extracted directly from Chen et al.\ data \cite{Chen_2020}, and subsequently used to produce the synthetic dataset.
}

\label{fig:self_validation_GITT}
\end{figure}%

\section{Inference using a GITT pulse-relax data}
\label{sec:validation_GITT}
We now apply the \ISDM to the experimental GITT pulse-relax data. Specifically, we consider the {GITT} charging cycle from which the diffusivity $D_{\rm GITT}(c)$ is inferred using the classical Sand equation approach.\footnote{
Note that $D_{\rm GITT}$ inferred from real data in this section is identical to the $\Dref$ used in \S\ref{sec:validation}.}
The result of applying the \ISDM to the same data is shown in Fig.\ref{fig:validation_GITT}. The \ISDM is in broad agreement with $D_{\rm GITT}$, exhibiting a similar shape, though slightly lower across most concentrations.

\begin{figure}[ht]
	\centering
	\includegraphics[width=.6\columnwidth]{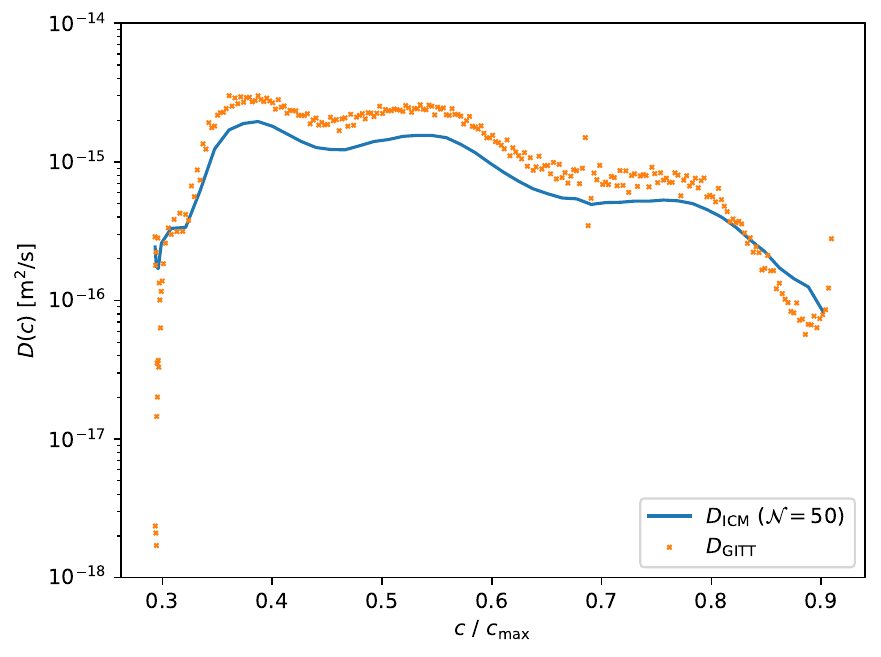}
	\caption{
		The diffusivity inferred from real GITT charging data \cite{Chen_2020} using \ISDM  with $\mathcal{N} = 50$ partitions.  The orange markers show $D_{\rm GITT}(c)$ obtained by applying the classical GITT analysis, i.e. using the Sand equation on each individual {GITT} pulse in the dataset. }
	\label{fig:validation_GITT}
\end{figure}%

In order to assess the quality of the inference we solve (\ref{eq:diffusioneq})-(\ref{modend}) parametrised by: (a) the diffusivity inferred by the \ISDM using the pulse-relax data,  and (b) $D_{\rm GITT}$, applying the experimental pulse-relax excitation, and
evaluating how well the predicted voltages match the experimental data using the metric (\ref{eq:R2_definition}). We find that
\begin{eqnarray}
	{\rm for \, \, the \, \, \ISDM:} \qquad R^2_{V~\mathrm{train}} = 0.501;\\
	{\rm for \, \, classical \, \, GITT:} \qquad R^2_{V~\mathrm{train}} = 0.470\,,
\end{eqnarray}
respectively.
That is, the \ISDM result accounts for just over half of the deviations in the data from the null model whereas the classical GITT method accounts for just under half.
The magnitude of the relative devation between the two inferred diffusivities $|1-D_{\rm \ISDM}(c)/D_{\rm GITT}(c)|$, averaged over the range of relevant concentrations $c/c_{\rm max} \in [0.3, 0.9]$ (see Fig. \ref{fig:validation_GITT}) is $34\%$.

Owing to the lack of a ground truth as to what the correct diffusivity is, it is
difficult to objectively assess which inferred diffusivity is more accurate. We shall return to this issue later in \S\ref{sec:comparison}; however, one might be tempted to speculate that the \ISDM result is more accurate on the grounds that it contains fewer embedded assumptions than the classical approach (there is no need to approximate diffusion in a sphere with that in an infinite slab, see assumption iv above). In any case, one clear trend is that the \ISDM result is markedly smoother than the classical one. This is appealing because the high frequency fluctuations in $\Dref$ are unlikely to derive from a real physical process.

\section{Inference using galvanostatic data}
\label{sec:validation_constant}

Thus far, our application of \ISDM has been limited to the pulse-relax current excitations designed to facilitate the classical inference approach. A significant strength of \ISDM lies in its versatility to handle \textit{any} current profile, provided that is does not violate assumptions 1--3 presented in \S\ref{sec:spm}. We now consider inference using data generated with constant C/10 current.

A vital component of the \ISDM methodology is the open circuit potential which, until now, has been derived from the {GITT} data. In order for the \ISDM to genuinely serve as a swift alternative to the classical approach, we require a means to infer the equilibrium potential, $U_{\rm eq}(c)$, without needing to resort to the slow-to-collect GITT data. Thus, we now perform inference with the \ISDM in two different scenarios: (a) using the open-circuit voltage (OCV) as derived from GITT data and then, (b) using a pseudo-open circuit voltage (pOCV) obtained from $C/20$ charge/discharge data\footnote{Here, we use the data set in \cite{Kendrick_C20_data} which was acquired from a half cell cathode and consists of alternating charge and discharge cycles at C/20, repeated 10 times. We estimate the pseudo open-circuit voltage as outlined in \ref{app:Ueq}, by averaging over the 10 charge/discharge cycles.}.
We extract the open-circuit voltage using a simple approach in which we assume that the measured cell voltage is made up of the open-circuit voltage and an Ohmic drop, which is equal in magnitude and opposite in sign between charge and discharge (since they occur at identical rates).
Thus, the Ohmic drop can be subtracted from the charging data and added onto the discharging data to arrive at the open-circuit voltage. The former test, (a), will allow us to assess the \ISDM's ability to infer diffusivity from fast-to-measure data whilst still leveraging a highly reliable characterisation of the open-circuit voltage. Successful inference of diffusivity in the latter setting, (b), would serve as strong evidence that the \ISDM can truly serve as a fast and accurate alternative to GITT. We show the results in Figure.\ref{fig:validation_constant}. The performance metrics are:
\begin{eqnarray}
	{\rm for \, \, the \, \, \ISDM\, \, and \, \,  OCV:} \qquad R^2_{V~\mathrm{train}} = 0.869\,;\\
	{\rm for \, \, the \, \, \ISDM\, \, and \, \,  pOCV:} \qquad R^2_{V~\mathrm{train}} = 0.878\,.
\end{eqnarray}

These performance metrics were computed consistently with the inference approach, i.e. we use the corresponding $U_{\rm eq}(c)$ that was used when inferring each $\hat{D}(c)$. The crucial conclusion is that although using the pseudo open-circuit voltage over that obtained from the GITT data has a small negative effect on the inference quality, with no appreciable impact on the inferred diffusivity.

\begin{figure}[ht]
	\centering\includegraphics[width=.6\columnwidth]{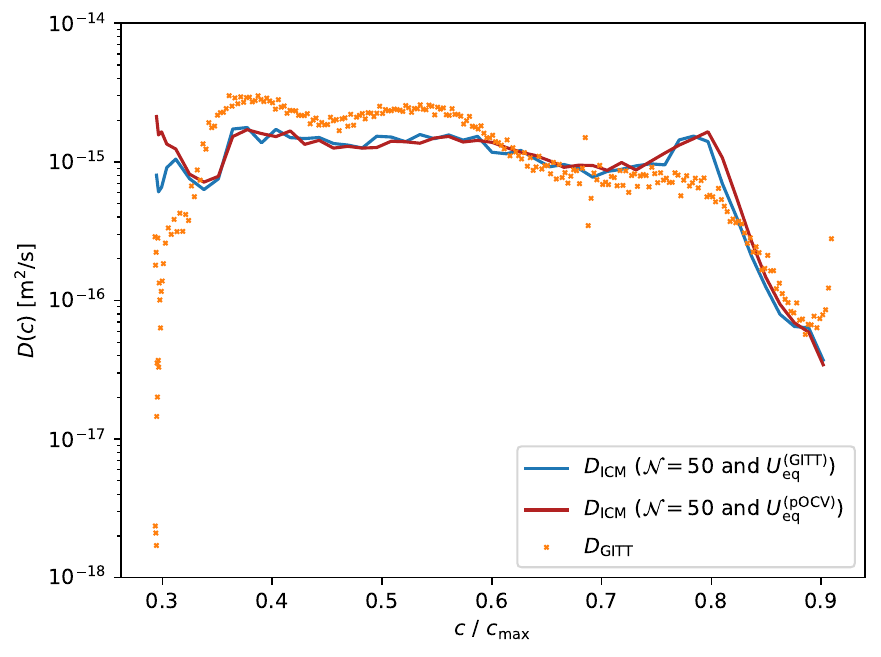}
	\caption{The diffusivity $D(c)$ inferred from \ISDM applied to real $C/10$ charging data from Ref.\cite{Chen_2020}, using $\mathcal{N} = 50$ partitions. The blue and red lines show $\hat{D}(c)$ inferred using $U_{\rm eq}(c)$ obtained from GITT and pOCV, respectively. The orange markers indicate the diffusivity obtained via the conventional GITT approach.
}
	\label{fig:validation_constant}
\end{figure}%

Before moving on we make a brief but relevant segue to assess the generalisability of the diffusivity inferred by entirely classical means, $D_{\rm GITT}$. To do this we solve the model (\ref{eq:diffusioneq})-(\ref{modend}) parameterised by $D_{\rm GITT}$ and excited by a galvanostatic charging current (cathode delithiation). The performance metric of this model in explaining the data is
\begin{equation} \label{badGITT}
{\rm for \, \, classical \, \, GITT \, \, and \, \,  OCV:} \qquad R^2_{V~\mathrm{test}} = 0.139\,.
\end{equation}
This value represents a modest improvement over the null model, which consists solely of the open-circuit voltage.
In Fig.\ref{fig:pred_vs_data} we show the predictions of the two inference approaches under the constant current ($C/10$) input discussed above. It is evident that the GITT-inferred diffusivity does not fully capture the dynamics seen in the experimental data.
\begin{figure}[ht]
	\centering\includegraphics[width=.6\columnwidth]{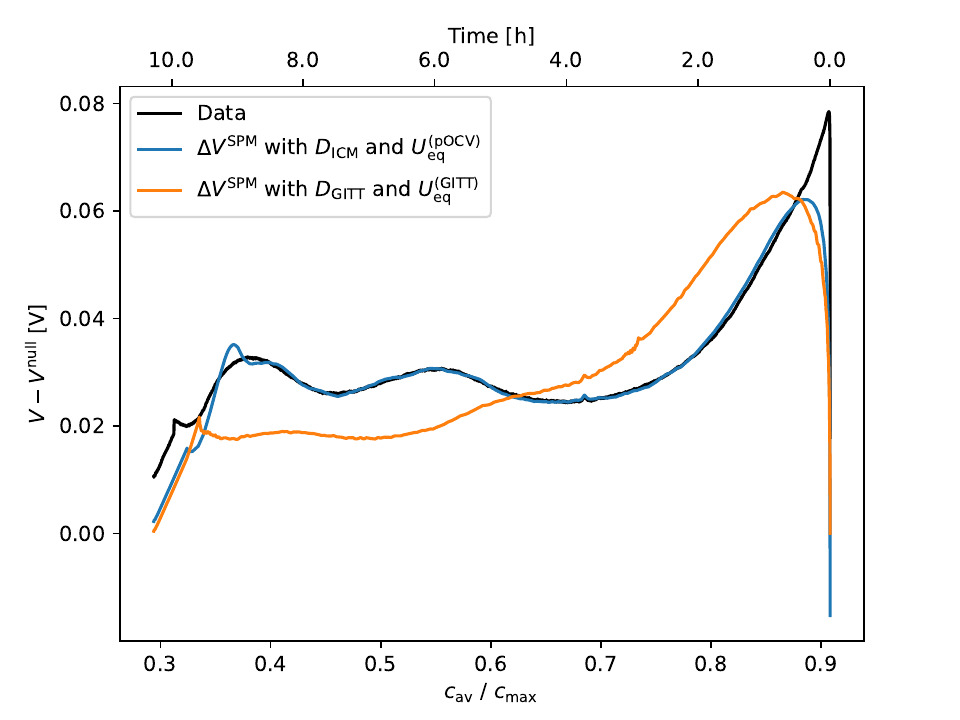}
	\caption{Comparison of SPM \eqref{eq:diffusioneq}-\eqref{modend} predictions using diffusivity inferred by \ISDM from constant current data (blue), and diffusivity inferred via Sand equation from GITT data (orange), under $C/10$ galvanostatic charging of a cathode half-cell. Actual measurements from Ref.\cite{Chen_2020} are shown for comparison in black. Here, the null model is selected as $V^{\rm null} = U_{\rm eq}^{\rm (GITT)}(c_{\rm av})$.
	}
	\label{fig:pred_vs_data}
\end{figure}%
Given that the inference model is inconsistent with the prediction setting, it is not surprising that the classically inferred diffusivity is less reliable; the metric (\ref{badGITT}) confirms that its generalisation performance is limited too.

In Table \ref{tab:Real_data_R2V}, we summarise the performance of the Sand equation and \ISDM approaches, evaluated using the two real datasets described above.
\begin{table}[h]
\centering
\begin{tabular}{ccc}
\toprule
Inference Method & Test set & $R^2_V$ \\
\midrule
Sand equation &  GITT & 0.470 \\
\ISDM & GITT & 0.402\\
Sand equation & C/10 GC & 0.139 \\
\ISDM & C/10 GC& 0.878\\
\bottomrule
\end{tabular}
\caption{
Performance metric $R^2_V$ for models inferred using either the Sand equation or the \ISDM approach, and evaluated on two experimental datasets: GITT and C/10 galvanostatic charge (GC). Models inferred using the Sand equation rely on GITT data and its corresponding OCV, while \ISDM models use GC data with OCV estimated from a separate pOCV dataset. In each case, the model is evaluated using the same OCV source as used during inference.}
\label{tab:Real_data_R2V}
\end{table}

\section{An even-handed comparison: \ISDM vs. GITT data and the Sand equation}
\label{sec:comparison}

Thus far, we have demonstrated i) self-consistency of the \ISDM, i.e. that it is capable to reproduce diffusivity provided that the data is consistent with the model embedded within the \ISDM; ii) that \ISDM outperforms classical approaches in explaining both pulse-relax GITT data and galvanostatic data.
However, to some extent, these latter results are unsurprising because the central premise of the \ISDM is to minimise deviations from observed data, see (\ref{optprob}).
What truly matters in practice is whether the \ISDM yields objectively more accurate diffusivities when deployed on genuine experimental data, or on synthetic data generated by solving a detailed, physically realistic model, such as the DFN.
To finally answer the question of whether \ISDM offers objective advantages we will now generate synthetic DFN data, which allows us to assess performance against a ground truth (i.e. the diffusivity used to parameterise the DFN model), and infer diffusivity using both the classical and \ISDM approaches. Furthermore, we will condition this test in favour of the classical approach by providing the Sand equation with slow-to-measure pulse-relax GITT data (generated using identical parameters to those described in \S\ref{expdat}), whereas the \ISDM will be given only a pseudo open-circuit voltage and fast-to-measure galvanostatic data (at rates of C/20 and C/10 respectively, as in the previous section). Dandeliion \cite{Korotkin_2021} was used to solve the full DFN model which we parametrised for an LG M50 NMC811 half-cell \cite{Chen_2020}. Links to the results of the simulations can be found in the footnote
\footnote{\label{ft:GIT}
	The Dandeliion run for {GITT} pulse-relaxation current can be accessed at:\\
	\href{https://simulation.dandeliion.com/legacy/simulation/?id=40751b48-307b-46fb-bcb9-5b3ab79b3b5c}{\texttt{ https://simulation.dandeliion.com/legacy/simulation/?id=40751b48-307b-46fb-bcb9-5b3ab79b3b5c}}.\\
	The Dandeliion run for charge and discharge with constant $C/20$ current can be accessed at:\\
	\href{https://simulation.dandeliion.com/legacy/simulation/?id=d91db0e5-e0bb-4391-936c-50c69f267c24}{\texttt{ https://simulation.dandeliion.com/legacy/simulation/?id=d91db0e5-e0bb-4391-936c-50c69f267c24}}\\
	The Dandeliion run for charging with constant $C/10$ current can be accessed at:\\
	\href{https://simulation.dandeliion.com/legacy/simulation/?id=3865ded1-60e1-439b-82b3-65a0ec8c5b0c}{\texttt{ https://simulation.dandeliion.com/legacy/simulation/?id=3865ded1-60e1-439b-82b3-65a0ec8c5b0c}}
}.
The pseudo open-circuit voltage was estimated using the same approach as in the previous section and as detailed in  \ref{app:Ueq}.

The comparison of the inferred diffusivities, along with the reference diffusivity $D_\textrm{ref}$ used to generate the synthetic data, is shown in Fig.\ref{fig:DFN_comparison}. The diffusivity performance metrics, as defined in (\ref{eq:R2_D}) and measuring objective accuracy of the diffusivity, are:
\begin{eqnarray}
	{\rm for \, \, the \, \, \ISDM\, \, and \, \,  pOCV:} \qquad R^2_D = 0.884\,;\\
	{\rm for \, \, the \, \, classical\, \, GITT \, \,  and \, \, OCV:} \qquad R^2_D = 0.780\,.
\end{eqnarray}
Thus, the \ISDM not only achieves a more accurate approximation of the diffusivity compared to classical approach in practice, but it also requires only 50 hours of data collection, in contrast to the 9.5 days necessary for the GITT data; around a five-fold reduction in measurement time.

\begin{figure}[ht]
\centering\includegraphics[width=.6\columnwidth]{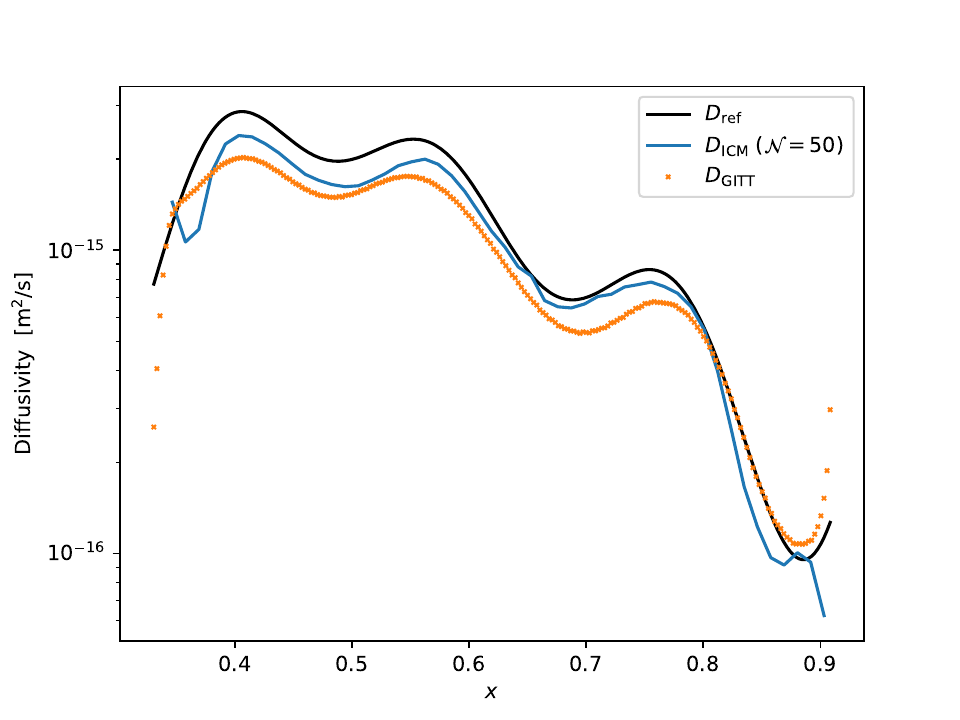}
\caption{Comparison of diffusivities inferred using \ISDM ($D_{\rm \ISDM}$, solid blue) with $\mathcal{N}=50$ partitions, and the traditional GITT approach ($D_{\rm GITT}$, orange markers), alongside the reference diffusivity (solid black) used to generate the DFN solution.
}
\label{fig:DFN_comparison}
\end{figure}%

Before closing this section, we summarise the performance of the Sand equation and \ISDM methods when evaluated on the synthetic datasets above, as shown in Table~\ref{tab:Synth_data_R2V}. Both approaches generalise reasonably well to unseen data, with the Sand equation performing marginally better in terms of test-set $R^2_V$. However, in terms of recovering the ground truth diffusivity, the \ISDM approach inferred from galvanostatic data performs better.
\begin{table}[h]
\centering
\begin{tabular}{lccll}
\toprule
Inference Method & Training set & Test set & $R^2_V$& $R^2_D$ \\
\midrule
\midrule
Reference &-- &GITT & 0.920 & 1\\
~&~ &C/10 GC& 0.903& \\
\midrule
Sand equation &  GITT & GITT & 0.758 & 0.780 \\
~&  ~& C/10 GC & 0.844\\
\midrule
\ISDM & GITT & GITT & 0.902 & 0.737 \\
& & C/10 GC& 0.799\\
\midrule
\ISDM & C/10 GC & C/10 GC & 0.981 & 0.884\\
& & GITT &  0.769\\
\bottomrule
\end{tabular}
\caption{
Performance metrics $R^2_V$ and $R^2_D$ for models inferred using either the Sand equation or the \ISDM approach, and evaluated on synthetic datasets: GITT and C/10 galvanostatic charge (GC). Models inferred from GITT data use the corresponding OCV, while the \ISDM model inferred from GC data uses OCV estimated from the separate pOCV dataset. In each case, the model is evaluated using the same OCV source as used during inference. The reference model denotes the SPM using the exact diffusivity and OCV functions used to generate the synthetic data.}
\label{tab:Synth_data_R2V}
\end{table}

\newpage

\section{Conclusions}
\label{sec:conclusions}

We have introduced a novel methodology for inferring the concentration-dependent solid-state diffusivity of battery active materials using experimentally measured voltage data, based on the single particle model (SPM). Our general approach, inference from a consistent model (\ISDM), relies on using the same underlying model for both inference and prediction, which, in this case, is the SPM.
The \ISDM approach offers the following advantages over the GITT:
\begin{enumerate}
\item It does not depend on the semi-infinite slab assumption inherent in the Sand equation; as we have shown in \ref{app:SlabtoSphere}, this assumption may lead to significant errors in the inferred diffusivities.
\item It does not require a carefully-controlled pulsed-current input and lengthy relaxation intervals. The \ISDM approach is highly versatile and can infer diffusivity from \textit{any} current excitation
that drives the concentration through the target range,
provided the experimental conditions conform to the SPM assumptions.
\item This adaptability to work with many data types allows significant savings in lab-time; as we have shown robust inference of the diffusivity is possible from fast-to-measure galvanostatic data. Moreover, it is likely that the \ISDM will perform similarly well on other, potentially even faster-to-measure data, provided its caveats are met.
\item We have demonstrated that, even when
provided solely with galvanostatic data,
the \ISDM more accurately infers diffusivity than the classical GITT and Sand equation approach. To make this concrete, in \S\ref{sec:comparison}, we showed that the \ISDM yields an objectively more accurate inference of the diffusivity with a five-fold reduction in data collection time over the classical method.
\item The \ISDM methodology can be extended in a systematic manner to account for complicating factors (e.g. electrolyte transport limitations, non-negligible overpotentials, Ohmic losses from electron conduction and charge transfer resistances) which may be encountered in practice. Such extensions come at the cost of model complexity and therefore a higher parameter count.
Thus, these parameters must either be well-characterised by supplementary experiments, or a more complex inference problem must be posed and solved to extract not only the diffusivity but also the additional parameters contemporaneously. This is likely to be a challenging task that remains open and might be approached by developing faster means of generating forward solutions using, e.g., ultra-fast surrogate models.
\end{enumerate}

It is worth noting that when applied to GITT data, \ISDM using the simple SPM formulation can be more susceptible to systematic drifts and formation effects than the Sand equation, which fits only the temporal evolution during each pulse and is thus largely robust against such offsets. Moreover, GITT data tend to contain a high degree of redundancy, making them suboptimal for the optimisation problem posed by \ISDM.

The most obvious application for our method is in inferring the diffusivity of an electrode at the beginning of life in order to make predictions of the device behaviour in the future. However, owing to the speed and agility of the technique it may prove useful in other ways. It is well-known that degradation, for example in the form of microscale intra-particle cracks, causes the effective diffusivity of the insertion material to reduce with cycle number. The \ISDM could be used in the field to update the model parameters from incoming data as the device ages, maintaining the predictive capability of the model. These parameter changes might also be used as an advanced indicator of approaching device failure or nonlinear ageing. Finally, although we have chosen to focus on solid-state diffusivity there are no conceptual hurdles to applying the technique elsewhere. In Li-ion batteries it could be used to infer electrolytic conductivity, transference number, or activation energies for the Arrhenius temperature dependence of the diffusivity. It may also prove useful in a variety of other energy capture and storage devices.

\section*{Acknowledgments}

We are grateful to Dr Edmund Dickinson (About:Energy Ltd) for his invaluable comments and suggestions for improvements during preparation. JF and EK were supported by the Faraday Institution Multi-Scale Modelling (MSM) project Grant number EP/S003053/1 (FIRG084). AEG, JB and JF were supported by a Faraday Institution Industrial Sprint project Grant number EP/S003053/1 (FIRG080).

\section*{Code Availability}
The code implementing \ISDM is available on GitHub at \href{https://github.com/gumrukcuoglu/ICM}{\texttt{gumrukcuoglu/ICM}}.

\appendix
\section{Validity of Slab approximation on a sphere}
\label{app:SlabtoSphere}

In this appendix, we discuss the validity of the slab approximation implicit in Sand equation to describe a spherical diffusion process. Similar analyses were presented in Ref.\cite{Nickol_2020} and in the Supplementary Information of Ref.\cite{PMID:37085556}. Our analysis focusses on quantifying the validity of $D(c)$ obtained using the Sand equation on an actual galvanostatic intermittent titration {technique (GITT)} experiment.

We consider the spherical diffusion equation \eqref{eq:diffusioneq} for constant diffusivity $D$, forced with a constant surface flux $j_0$, and starting with an initial homogeneous concentration $c_0$. We rescale concentration $c$, distance $r$ and time $t$ with initial concentration $c_0$, particle radius $R$ and diffusion timescale $t_d$, respectively. That is,
\begin{equation}
\tilde{c}(t,r) = \frac{c(t,r)}{c_0}\,,\qquad x = \frac{r}{R}\,,\qquad
\tau = \frac{t}{t_d} = \frac{D\,t}{R^2}\,.
\label{eq:rescaling}
\end{equation}
With these redefinitions, Eq.\eqref{eq:diffusioneq} becomes
\begin{equation}
\frac{\partial \tilde{c}(\tau,x)}{\partial \tau} = \frac{\partial^2\tilde{c}(\tau, x)}{\partial x^2} +\frac{2}{x}\,  \frac{\partial\tilde{c}(\tau, x)}{\partial x}
\,,
\label{eq:diffusioneq_redef}
\end{equation}
subject to \eqref{eq:midprob}, which under \eqref{eq:rescaling}, become:
\begin{equation}
\tilde{c}(0,x)=1\,, \qquad
\left.\frac{\partial \tilde{c}(\tau,x)}{\partial x}\right\vert_{x=0} = 0 \,,\qquad
\left.\frac{\partial \tilde{c}(\tau,x)}{\partial x}\right\vert_{x=1} = -\delta \,.
\end{equation}
Here, we defined
\begin{equation}
\delta = \frac{j_0\,R}{c_0D\,F}\,.
\end{equation}
The solution to Eq.\eqref{eq:diffusioneq_redef} is known analytically \cite{Carslaw1959-to, SUBRAMANIAN2001385},
\begin{equation}
\tilde{c}(\tau,x) = 1 - \delta\,\left[3\,\tau + \frac{ 5\,x^2 - 3}{10} - \frac{2}{x}\,\sum_{n=1}^{\infty}\frac{\sin(\alpha_n\,x)}{\alpha_n^2\,\sin(\alpha_n)}\,{\rm e}^{-\alpha_n^2\tau}\right]\,,
\end{equation}
where $\alpha_n$ is the $n$--th positive root of the transcendental equation $\alpha = \tan\alpha $.

On the surface ($x=1$), this solution reduces to:
\begin{equation}
\tilde{c}(\tau,1) = 1 - \delta\,\left[3\,\tau + \frac{ 1}{5} -2\,\sum_{n=1}^{\infty}\frac{{\rm e}^{-\alpha_n^2\tau}}{\alpha_n^2}\right]\,.
\label{eq:fullsol}
\end{equation}

At $\tau=0$, the sum evaluates to $\sum_{n=1}^\infty \alpha_n^{-2} = 1/10$ \cite{Liron71}, consistent with the initial condition $\tilde{c}(0,1)=1$.

We will now discuss the asymptotic behaviour of this expression.
One way to determine the behaviour of the infinite sum is to approximate it by an integral. We first observe that for large $n$, the recursion approximates to $\alpha_{n+1} \approx \alpha_n +\pi$. Extending this relation, we have
\begin{equation}
\alpha_n \approx \alpha_1 +(n-1)\pi\,,
\end{equation}
which is accurate at 1\% level for low $n$, but becomes more precise for larger values. This motivates the approximation
\begin{equation}
S = \sum_{n=1}^\infty \frac{{\rm e}^{-\alpha_n^2\tau}}{\alpha_n^2} \approx \sum_{n=1}^\infty s(n) \,,
\label{eq:infinite_sum}
\end{equation}
where we defined
\begin{equation}
s(n) = \frac{{\rm e}^{-[\alpha_1+(n-1)\pi]^2\tau}}{[\alpha_1+(n-1)\pi]^2}\,.
\end{equation}
Next, we approximate the infinite sum \eqref{eq:infinite_sum} with the Euler-Maclaurin formula that allows us to convert the discrete sum to a continuous integral:
\begin{equation}
S=\sum_{n=1}^\infty s(n) \approx \int_1^\infty s(x) dx + \frac{s(\infty)+s(1)}{2} + \sum_{k=1}^{\infty}\frac{B_{2k}}{(2k)!}\,\left(s^{(2k-1)}(\infty) - s^{(2k-1)}(1)\right)\,,
\label{eq:bernouilli}
\end{equation}
where $B_{2k}$ are the even Bernoulli numbers and $s^{(2k-1)}$ is the $k$-th odd derivative of $s(n)$. Since we are only interested in small $\tau$ approximation, we only keep terms up to linear order in $\tau$ in the Taylor expansion. The first term of \eqref{eq:bernouilli} is straightforward to evaluate, and yields:
\begin{equation}
\int_1^\infty s(x)dx = \frac{{\rm e}^{-\alpha_1^2\tau}}{\alpha_1\pi}- \sqrt{\frac{\tau}{\pi}}\,{\rm erfc}\left(\alpha_1 \sqrt{\tau}\right) = \frac{1}{\alpha_1\pi} - \sqrt{\frac{\tau}{\pi}} +\mathcal{O}(\tau)
\label{eq:integral_part}
\end{equation}
Regarding the correction terms in Eq.\eqref{eq:bernouilli}, we note that $s(n)$ and all its odd derivatives vanish as $n\to\infty$. The only remaining contributions come from the function and its derivatives at $n=1$.
The infinite sum of odd derivatives at $n=1$ are problematic. Around $k=12$th term, the sum starts to grow rapidly. On the other hand, in the small $\tau$ expansion, all contributions from odd derivatives give contributions at order $\tau^0$, $\tau^2$ and higher. Since we are mainly interested in the behaviour of the $\tau^{1/2}$ term, we will assume that these terms can be resummed to a well-behaving form. In the end we are left with a constant term and the $\sqrt{\tau}$ term obtained in \eqref{eq:integral_part}. We can fix the constant contribution by using the finite limit for $\tau\to0$ and obtain
\begin{equation}
S \approx \frac1{10}-\sqrt{\frac{\tau}{\pi}} +\mathcal{O}(\tau)\,.
\end{equation}
That is, for small $\tau$, the solution to the spherical diffusion equation is  approximated to
\begin{equation}
\frac{\tilde{c}(\tau,1)-1}{\delta} = -2\,\sqrt{\frac{\tau}{\pi}} +\mathcal{O}(\tau)\,.
\label{eq:early_approx}
\end{equation}
Written in dimensionful quantities, the surface concentration is:
\begin{equation}
c_{\rm surf} =
c_0 +\frac{j_0\,R}{D\,F}\left(-\frac{2}{\sqrt{\pi}}\,\sqrt{\frac{D\,t}{R^2}}+ \mathcal{O}\left(\frac{D\,t}{R^2}\right)\right) \approx c_0 - 2\,\frac{j_0}{F}\,\sqrt{\frac{t}{D\,\pi}}\,, \quad {(t \ll R^2/D)}
\label{eq:fishy_equation}
\end{equation}
Thus, for $t \ll R^2/D$, the spherical diffusion equation has the same surface solution as the diffusion equation for a semi-infinite slab \cite{Weppner1977} which forms the basis of the Sand equation. Although Eq.\eqref{eq:fishy_equation} suggests that GITT may be applicable for short pulses, a quantitative threshold for the validity of this approximation is still required.

Moreover, at relatively large $\tau$, the infinite sum in Eq.\eqref{eq:fullsol} decays exponentially, and we are left with
\begin{equation}
\frac{\tilde{c}(\tau,1)-1}{\delta} \approx -\frac15-3\,\tau\,,
\label{eq:late_approx}
\end{equation}
or
\begin{equation}
c_{\rm surf} \approx  c_0 - \frac{j_0R}{D\,F}\left(\frac15+\frac{3\,t\,D}{R^2}\right)\,,
\qquad (t\gtrsim R^2/D)\,.
\label{eq:useless_equation}
\end{equation}
This expression is particularly important since the surface concentration depends linearly on $t$ and the slope is independent of $D$. Therefore if one employs the GITT in this regime, the inferred diffusion constants will be arbitrary. Again, we need to use numerical methods to determine precisely the time $\tau = t\,D/R^2$ where this behaviour becomes relevant.

In Figure~\ref{fig:approx_comparison}, we show a comparison of the full solution and demonstrate the validity of \eqref{eq:early_approx} and \eqref{eq:late_approx} in the corresponding regimes.
\begin{figure}[ht]
\centering
\includegraphics[width=0.6\columnwidth]{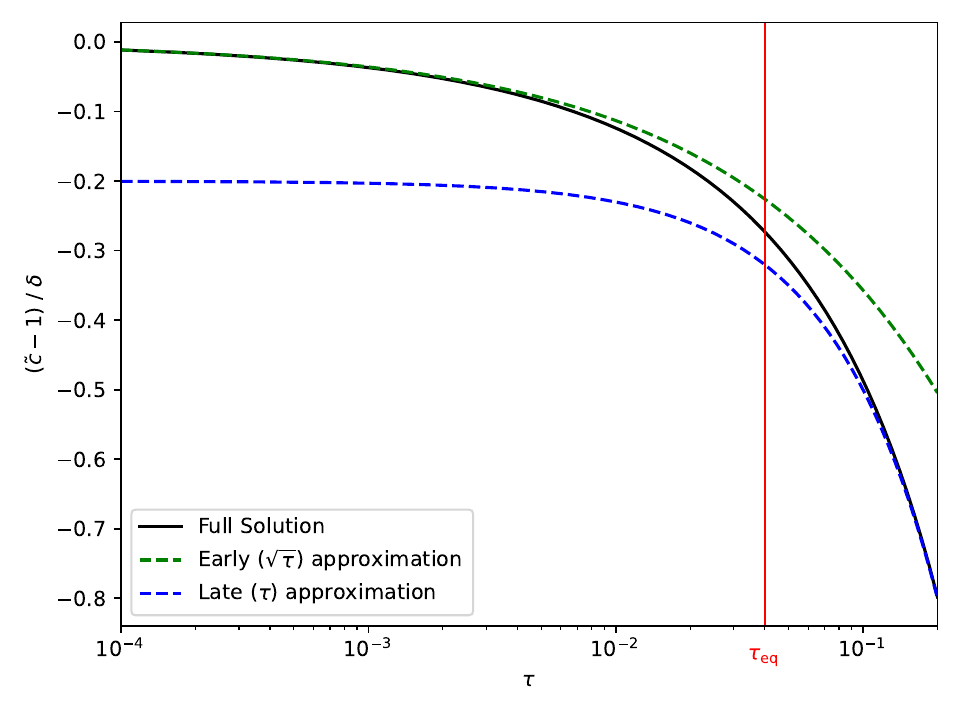}
\caption{Comparison of the full solution (solid black), the early $\sqrt{\tau}$ approximation (green dashed) and the late linear behaviour (blue dashed). The crossover point (red) marks where neither approximation dominates.}
\label{fig:approx_comparison}
\end{figure}

To quantify the accuracy of the approximations, we calculate the relative error by
\begin{equation}
{\rm Error[\%]} = 100\times\frac{\tilde{c}-\tilde{c}_{\rm approx}}{\tilde{c}-1}\,,
\end{equation}
where $\tilde{c}$ is the exact solution while $\tilde{c}_{\rm approx}$ is the approximation. The time dependence of the relative errors are presented in Figure \ref{fig:approx_pererr}.
\begin{figure}[ht]
\centering
\includegraphics[width=0.6\columnwidth]{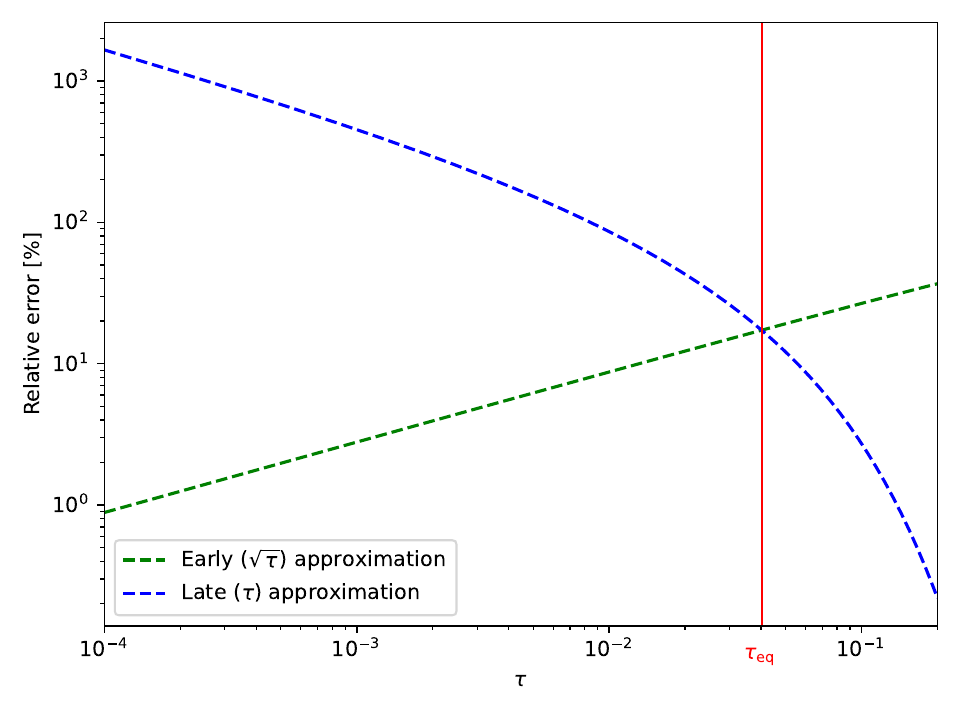}
\caption{
Relative error between full solution and early/late-time approximations. The crossover point ($\tau_{\rm eq}$) marks where neither approximation dominates.}
\label{fig:approx_pererr}
\end{figure}%
We see that the square-root solution \eqref{eq:early_approx} is valid within 5\% for $\tau<0.0032$, 7.5\% for $\tau <0.0073$ and 10\% for $\tau <0.0132$.

On the other hand, we see that
the relative error of the late solution \eqref{eq:late_approx} catches up with that of the early approximation \eqref{eq:early_approx} at $\tau_{\rm eq} = 0.0402$, with both errors at $17.25\%$. For $\tau >\tau_{\rm eq}$, the linear approximation rapidly improves and the Sand equation is no longer applicable. \footnote{A typical value quoted in the literature for the $5\%$ accuracy of the linear solution is $\tau > 1.27$ \cite{Nickol_2020, PMID:37085556}. This actually corresponds to the accuracy associated with approximating $(\tilde{c}-1)/\delta\approx -3\,\tau$, i.e. without the constant term. Instead, including the intercept $(\tilde{c}-1)/\delta \approx -1/5 -3\,\tau$, the linear behaviour becomes valid at a much earlier time, reaching $5\%$ accuracy at $\tau>0.0783$. }

In Figure~\ref{fig:approx_validity} we demonstrate the effect of approximating the sphere as a semi-infinite slab using the GITT data for cathode delithiation from Ref.\cite{Chen_2020}.  Here $\tau = t\,D/R^2$ is computed for each GITT pulse of duration $t$, with $D$ inferred from the Sand equation. In this dataset, the value of $\tau$ is consistently above the 5\% cutoff, exceeding the 7.5\% threshold over roughly half the range. Overall, the Sand equation remains valid within 10\% accuracy.
\begin{figure}[ht]
\centering
\includegraphics[width=0.6\columnwidth]{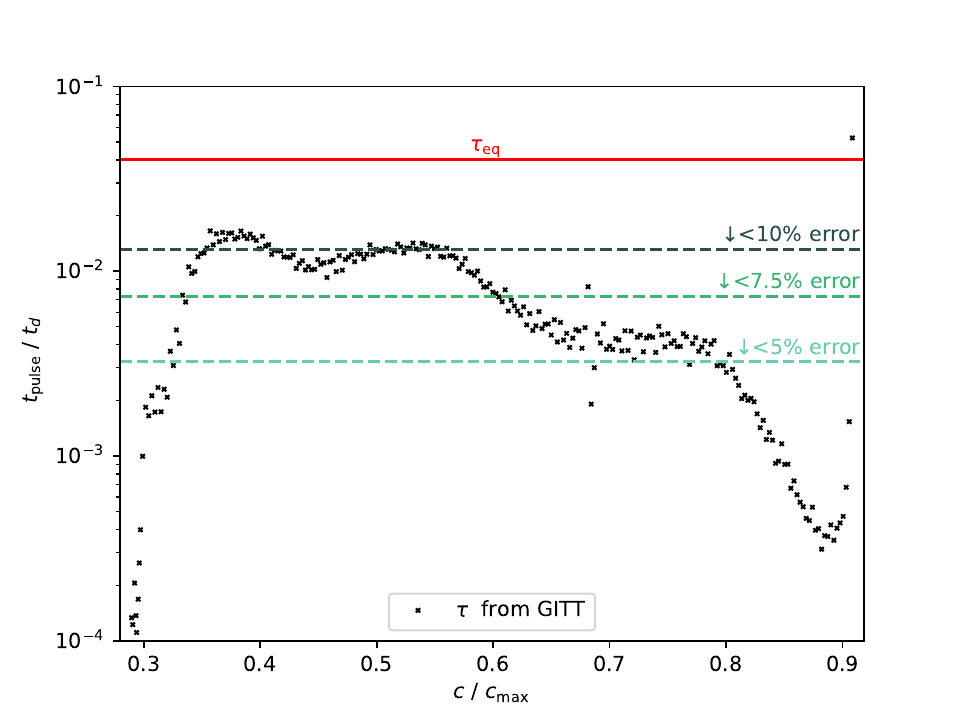}
\caption{Plot of the $\tau$ parameter for Chen et al \cite{Chen_2020} cathode delithiation data. The dashed lines denote the 5\%, 7.5\% and 10\% accuracy lines, respectively. The solid red line corresponds to the critical $\tau_{\rm eq}$ value above which, one cannot infer diffusivity from the Sand equation.}
\label{fig:approx_validity}
\end{figure}%

This example underlines the fact that, while the Sand equation is formally valid below the breakdown threshold $\tau_{\rm eq}$, it may still incur approximation errors in the surface concentration as large as $17\%$.
Given a relative error $\Delta c_s/c_s$ in the surface concentration, Eq.\eqref{eq:fishy_equation} implies that the relative error in diffusivity is
\begin{equation}
\frac{\Delta D}{D} = -2\,\frac{\Delta c_s}{c_s}\,.
\label{eq:DeltaD}
\end{equation}
Since the early solution \eqref{eq:fishy_equation} consistently underestimates the drop in concentration, this leads to an overestimation of diffusivity larger by a factor of 2. For instance, a $5\%$ error in $\tilde{c}$ would translate into a $10\%$ error in $D$.

\section{Optimisation strategies for inferring diffusivity}
\label{app:optapp}

Inferring the concentration-dependent diffusivity function $D(c)$ from current–voltage data is a computationally challenging task. In this work, we parameterise $D(c)$ with $\mathcal{N}$ parameters, each corresponding to the knots of a linear interpolation function on a stoichiometry (or concentration) grid. The resulting optimisation problem requires a search in $\mathcal{N}$-dimensional parameter space, an increasingly difficult problem as $\mathcal{N}$ grows.
Moreover, the problem is further complicated because the objective function depends on diffusivity through the solution of a partial differential equation. This is particularly challenging when time series data span weeks and require fine temporal resolution to resolve abrupt transients (e.g., at the onset and termination of GITT pulses). These computational challenges are avoided when using the Sand equation with GITT data, since the inference reduces to evaluating an analytical approximation rather than solving the full diffusion PDE. As we discussed in the Introduction, this involves destructive disassembly of the full cell to isolate the electrode, as well as a lengthy GITT experiment.

A naive approach would attempt to minimise the global loss function directly in this high-dimensional space, but such an approach is computationally costly and may suffer from convergence issues, particularly when no reliable initial estimate is available. The difficulty is compounded by the stiffness of the diffusion equation and the potential for multiple local minima. There are of course, many possible approaches to lighten the load of the optimisation.
One possibility, when some prior characterisation is available (e.g., for similar chemistries or devices), is to initialise the optimisation using a lower-dimensional parameterisation informed by previous measurements.

Another alternative would be to use the adjoint sensitivity method \cite{cacuci2003sensitivity},
which facilitates efficient use of gradient-based algorithms, potentially accelerating the search for $\hat{\theta}$ (see e.g. Ref.\cite{KUZHIYIL2025125221} for an application to the battery context). Nonetheless, for the present study, since the major focus is validation and benchmarking against a {GITT} experiment with a very large number of data points, a global parameterisation with adjoint methods proves to be computationally expensive and can become unstable due to the stiffness of the diffusion equation.

In this work, we instead adopt a two-stage optimisation strategy, allowing us to analyse individual data segments with localised parameters, providing a more practical application of \ISDM for the present dataset.

In the first stage, the problem is decomposed into a sequence of $\mathcal{N}$ independent scalar optimisation problems by partitioning the dataset into segments over which the diffusivity can be locally approximated as constant. This decomposition allows for efficient estimation of diffusivity values at each partition through simple grid search methods, yielding an initial guess for the parameter vector.

In the second stage, these initial estimates are refined via a global optimisation that incorporates the full time-dependent behaviour of the system, ensuring that the final inferred diffusivity profile is consistent with the data across all partitions. This staged approach provides both computational tractability and robustness, while retaining the flexibility to adapt to different experimental protocols, such as GITT or galvanostatic cycling.

In the following we present the details of the optimisation strategy.

\subsection{Choosing partitions}

We first segregate the time series data into $\mathcal{N}$ partitions. To each partition $\alpha$, we assign the subset of data:
\begin{equation}
	\mathcal{D}_\alpha = \left\{t_{\alpha i}, I_{\alpha i}, V_{\alpha i} \right\}_{i=1}^{N_\alpha}\,,
\end{equation}
where $N_\alpha$ is the number of data points in partition $\alpha \in \{1,\dots,\mathcal{N}\}$

For the success of the first (local) optimisation stage, the selection of the number and locations of the partitions is a crucial step.
The optimal choice depends on both the form of the input current and the temporal density of measurements, such that each partition contains sufficient information on the lithium diffusion process.
For instance, a typical {GITT} experiment consists mostly of stabilisation data, where the potential does not noticeably change; these data points contain limited information once the system is close to equilibrium. Thus, for {GITT} data, a natural choice is to select the partitions that each contain at least one pulse-relaxation cycle, thereby democratically distributing the available diffusion information.

In cases where there is no obvious choice for partitioning, e.g. experiments with continuous constant current input (see Section \ref{sec:validation_constant}), one option is to divide the data such that each partition contains an equal number of \textit{useful} data points. In the context of galvanostatic charging at constant rate, this is simply achieved by using equal time intervals, as the transferred charge per partition is uniform. For more complex or variable input profiles, such as drive-cycle data, where the information density across the time-series may vary substantially, one can define a ‘useful’ point as one where the voltage change $\Delta V$ exceeds a pre-set threshold $\Delta V_{\rm min}$, and ensure a more balanced distribution of information across partitions.

Once the partitions are selected, we can assign Li ion concentration values to each. For this, we use the space and time averaged concentration for the partition $\hat{c}_\alpha$, as
\begin{equation}
	\hat{c}_\alpha = \frac{3}{T_\alpha\,R^3}\int_0^R \,\int_{t_{\alpha1}}^{t_{\alpha1}+T_\alpha}\,r^2 c(t, r ; \hat{\theta}_\alpha) \,dt\,dr\,,
\end{equation}
which, on leveraging (\ref{eq:diffusioneq})-(\ref{eq:endprob}),
can be rewritten more conveniently (in the sense that it can be precomputed because it is independent of the decision variables) as
\begin{equation}
	\hat{c}_\alpha = \frac{1}{T_\alpha} \int_{t_{\alpha1}}^{t_{\alpha1}+T_\alpha} c_{\rm av}(t) \,dt\,,
\end{equation}
where $T_\alpha$ is the duration occupied by partition $\alpha$ and $c_{\rm av}(t)$ is the volume-averaged concentration across the particle given by
\begin{equation}
c_{\rm av}(t) = \left( \frac{3}{R^3} \int_0^R r^2 c_0(r) \,dr \mp \frac{3}{4\,\pi\,R^3n\, F} \int_0^t I(s) \, ds \right)\,.
\label{eq:cav_def}
\end{equation}

Any inferred parameter within partition $\alpha$ will be assigned a concentration value $\hat{c}_\alpha$. In particular, we parameterise the diffusivity function  $D=D(c; \theta)$ as a piecewise linear function, where $\theta = (\theta_1, \theta_2, \dots, \theta_\mathcal{N})$ is the parameter vector. The knots of the interpolation function are located at $(\hat{c}_\alpha, \theta_\alpha)$. The locations of the concentration values are data and partitioning dependent.

\subsection{Stage one: local optimisation over data partitions}

For the first stage of the optimisation, we consider $\mathcal{N}$ independent inference problems.
We evolve the diffusion equation for each partition sequentially (by taking the final concentration distribution for one partition as the initial condition for the next) assuming a series of Fickian diffusion processes with constant diffusivity $\theta_\alpha$ in each partition, and solve the corresponding optimisation problem; i.e. by defining the local loss in partition $\alpha$ as
\begin{equation}
	\mathcal{L}_\alpha(\theta_\alpha) = \frac{1}{N_\alpha}\sum_{i=1}^{N_\alpha} \left[V_{\alpha i} - U_{\rm eq}(c(t_{\alpha i},R ; \theta_\alpha))\right]^2\,,
\end{equation}
we estimate the constant diffusivity implied by the data over each partition $\theta_\alpha$ as
\begin{equation}
	\hat{\theta}_{\alpha} = \underset{\theta_{\alpha}}{\rm argmin} \,\mathcal{L}_{\alpha}(\theta_{\alpha})\,.
\end{equation}
Solution of each optimisation problem is relatively straightforward because it contains a single scalar decision variable, $\hat{\theta}_\alpha$; thus a grid search approach is feasible.

For the first stage of the optimisation, we carry out grid search for $D \in \left( 10^{-17},10^{-13}\right)$ m$^2$/s in log-scale over each of the $\mathcal{N}$ partitions sequentially. The grid is iteratively refined in the neighbourhood of the optimum until the solution converges to within 0.1\% accuracy. If no minimum is found within the starting range, the number of grid points is expanded by a factor of $10$, and the range is extended by a factor of $10$ on both sides. As a result, we obtain $\mathcal{N}$ pairs of $(\hat{c}_\alpha, \hat{\theta}_\alpha)$.
These pairs serve as informed guesses for the positions and values of the diffusivity at the interpolation knots. Thus, they provide a good initial candidate from which we can begin the search for $\hat{\theta}$ that solves the global optimisation problem (\ref{optprob}).

For scenarios in which the concentration gradients within the particle are minimal, e.g. a GITT experiment with low constant current pulses interspersed with long and frequent relaxation periods, the result from step one is likely to be a very good approximation of $\hat{\theta}$. By contrast, for scenarios where the data is measured with less sedate current excitations, such that there are significant concentration gradients within the particles, the second optimisation step is required to yield satisfactory results.

Note that this strategy of breaking down the inference of the concentration-dependent diffusivity into a series of single-decision-variable optimisation problems has a clear parallel to the classical GITT process. In fact, the pulsed current excitation used in classical GITT is designed to facilitate the familiar piecewise approach to the optimisation. However, for a general experimental protocol, it only acts as a first approximation of a more complete optimisation strategy.

\subsection{Stage Two: Global Refinement}

In the second optimisation stage, the initial estimates from the first stage are refined by solving the global optimisation problem \eqref{optprob} using a numerical gradient-based algorithm.

While the first stage provides a good initialisation, it treats each partition independently and assumes that the diffusivity within each partition can be assigned to a single representative concentration. In practice, however, each partition samples a range of concentrations, particularly when the current is large, the partition duration is long, or diffusion is slow. As a consequence, the effect of changing diffusivity at one concentration can propagate across multiple partitions. The global optimisation stage accounts for these interdependencies, ensuring that the final diffusivity profile remains consistent with the entire dataset and the full time-dependent evolution of concentration.

For general current protocols, where concentration overlaps between partitions are uncontrolled and potentially widespread, we employ a standard gradient descent algorithm. The global loss function \eqref{eq:totalloss} is minimised by numerically estimating the gradients with respect to all parameters $\theta_\alpha$, and updating the full parameter vector simultaneously. This fully coupled optimisation is employed for the galvanostatic charging data presented in the main text.

By contrast, for protocols where concentration overlaps remain confined to neighbouring partitions, a simplification is possible. In GITT experiments, for example, the relatively low pulse currents combined with long relaxation intervals limit the development of strong intra-particle concentration gradients. That is, the concentration remains close to its average within each partition.
In such cases, we adopt a localised cyclic gradient descent scheme, where each parameter $\theta_\alpha$ is updated one by one, while considering data from only its $2\beta$ nearest neighbours. This substantially reduces the number of forward PDE evaluations, which is especially advantageous in GITT regimes, where the diffusion equation is stiff and costly to solve.
Instead of using the global loss function directly, we define a quasi-global loss function for each partition,
\begin{equation}
	\mathcal{L}_\alpha^{(\beta)}(\theta_\alpha) = \frac{1}{\sum_{i=\alpha-\beta}^{\alpha+\beta} N_i} \,\sum_{i=\alpha-\beta}^{\alpha+\beta} \,\sum_{j=1}^{N_i}\left[V_{i,j} - U_{\rm eq}\left(c(t_{i,j}, R; \theta_\alpha)\right)\right]^2\,,
	\label{eq:quasi-global_loss}
\end{equation}
where $\beta$ denotes the number of neighbouring partitions included on either side of $\alpha$. For the synthetic GITT data used in this work, we set $\beta=2$, so that each local update includes information from 5 adjacent partitions. The gradient of
$\mathcal{L}_\alpha^{(\beta)}(\theta_\alpha)$
with respect to $\theta_\alpha$ is estimated numerically and used to update $\theta_\alpha$. The procedure is repeated cyclically over all partitions until the global loss $\mathcal{L}(\theta)$ converges, yielding the final inferred diffusivity $\hat{D}(c) = D(c;\hat\theta)$.

\subsection{Pseudocode}

Here we present the pseudocode for the optimisation algorithm.
\begin{algorithmic}
\Procedure{InferDiffusivity}{Data, $\mathcal{N}$, \texttt{N\_iter}, $\beta$}
    \Comment{Stage 1: Local optimisation in partitions}
    \State Divide \texttt{Data} into $\mathcal{N}$ partitions: $\{(t_\alpha, I_\alpha, V_\alpha)\}$
    \For{each partition $\alpha = 1$ to $\mathcal{N}$}
        \State Compute $\hat{c}_\alpha$ as average Li concentration in partition $\alpha$
        \State Define local loss $\mathcal{L}_\alpha(\theta_\alpha)$ using $(t_\alpha, V_\alpha)$
        \State Perform grid search to minimise $\mathcal{L}_\alpha$ and find $\hat{\theta}_\alpha$
        \State Store knot pair $(\hat{c}_\alpha, \hat{\theta}_\alpha)$
    \EndFor
    \State Construct initial $\hat{D}(c)$ as piecewise linear through knots $(\hat{c}_\alpha, \hat{\theta}_\alpha)$

    \Comment{Stage 2: Global refinement}
    \If{concentration overlap is localised (e.g. GITT)}
        \For{iteration $= 1$ to \texttt{N\_iter}}
            \For{each partition $\alpha$}
                \State Compute gradient $\partial \mathcal{L}_\alpha^{(\beta)} / \partial \theta_\alpha$ numerically
                \State Update $\theta_\alpha$ using gradient descent
            \EndFor
        \EndFor
    \Else \Comment{e.g. continuous C/10 current}
        \For{iteration $= 1$ to \texttt{N\_iter}}
            \State Compute $\nabla_\theta \mathcal{L}$ numerically for all $\theta_\alpha$
            \State Update all $\theta_\alpha$ simultaneously using gradient descent
        \EndFor
    \EndIf

    \State \Return final diffusivity profile $\hat{D}(c)$
\EndProcedure
\end{algorithmic}

\section{Determination of equilibrium potential using pOCV}
\label{app:Ueq}

The efficacy of \ISDM, like any inference technique, hinges on accurately determining the equilibrium potential $U_{\rm eq}(c)$. In particular, if the output voltage differs from the correct $U_{\rm eq}$ once the system has relaxed, the inference process becomes unreliable, yielding biased or meaningless estimates.

When we are presented with {GITT} data, it is straightforward to reconstruct the $U_{\rm eq}(c)$ function from the voltages at the relaxation points. However, to make use of the flexibility of \ISDM, it is crucial to find alternative approaches to determining $U_{\rm eq}(c)$  without resorting to lengthy experimental techniques.

In this appendix, we outline the determination of the $U_{\rm eq}^{(pOCV)}(c)$ function used in Sections \ref{sec:validation_constant} and \ref{sec:comparison}.
In order to construct the $U_{\rm eq}$, we first note that
\begin{equation}
V_{\rm measured} = U_{\rm eq}(c_{\rm surf}) \pm |\eta|\,,
\end{equation}
where $\eta$ is the magnitude of the reaction overpotential; the sign is positive for charging and negative for discharging. Assuming a small enough current such that the charge is quickly dissipated across the particle, we can estimate the equilibrium potential as:
\begin{equation}
U_{\rm eq} \approx \frac12\,\left(V_{\rm charge} + V_{\rm discharge}\right)\,,
\end{equation}
such that the effect of the overpotential is cancelled. For the dataset used in Sec.\ref{sec:validation_constant}, containing 10 charge-discharge cycles, we average over all 10 cycles,
and show the estimated $U_{\rm eq}^{\rm (pOCV)}$ and the corresponding $U_{\rm eq}^{\rm (GITT)}$ in Fig.\ref{fig:Ueq}. The mean squared error between the two functions is $7\times 10^{-6}$ within the overlapping stoichiometry range.
\begin{figure}[ht]
\centering
\includegraphics[width=0.6\columnwidth]{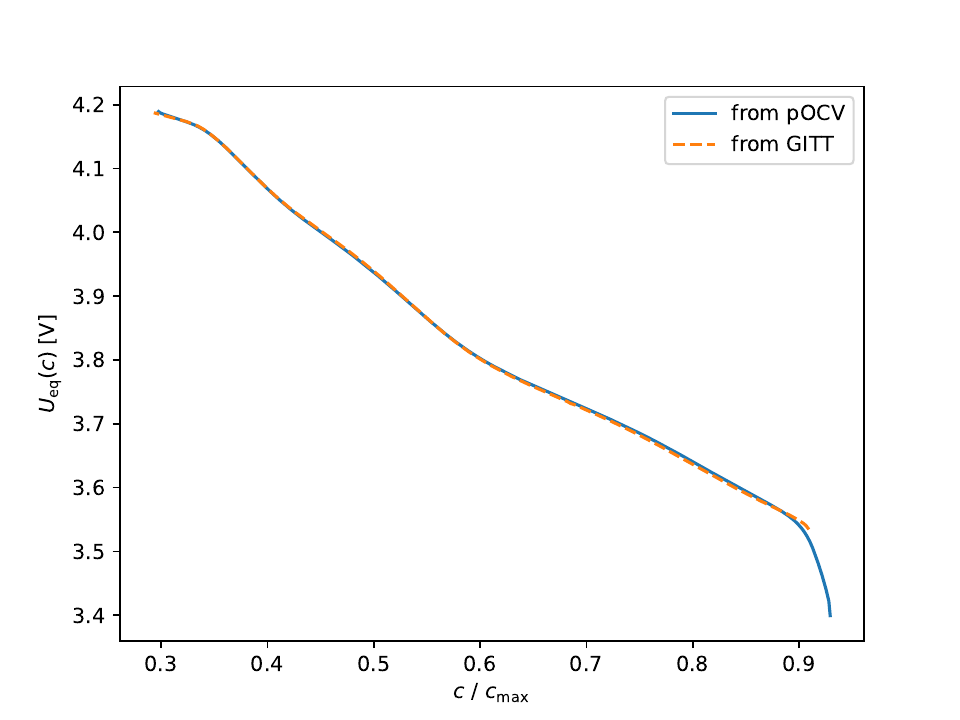}
\caption{The equilibrium potential
estimated from pOCV data collected under constant C/20 current, compared to the one obtained directly from {GITT} data.}

\label{fig:Ueq}
\end{figure}

\bibliographystyle{elsarticle-num}
\bibliography{NeuralDiffusivity}

\begin{thebibliography}{10}
\expandafter\ifx\csname url\endcsname\relax
  \def\url#1{\texttt{#1}}\fi
\expandafter\ifx\csname urlprefix\endcsname\relax\def\urlprefix{URL }\fi
\expandafter\ifx\csname href\endcsname\relax
  \def\href#1#2{#2} \def\path#1{#1}\fi

\bibitem{doyle93}
M.~Doyle, T.~F. Fuller, J.~Newman, Modeling of galvanostatic charge and
  discharge of the lithium/polymer/insertion cell, Journal of the
  Electrochemical Society 140~(6) (1993) 1526--1533.

\bibitem{fuller94}
T.~F. Fuller, M.~Doyle, J.~Newman, Simulation and optimization of the dual
  lithium ion insertion cell, Journal of the Electrochemical Society 141~(1)
  (1994) 1--10.

\bibitem{fuller94b}
T.~F. Fuller, M.~Doyle, J.~Newman, Relaxation phenomena in
  lithium-ion-insertion cells, Journal of the Electrochemical Society 141~(4)
  (1994) 982--990.

\bibitem{1994FuDoNe}
T.~F. Fuller, M.~Doyle, J.~Newman, Simulation and optimization of the dual
  lithium ion insertion cell, J. Electrochem. Soc. 141~(1) (1994) 1--10.
\newblock \href {https://doi.org/10.1149/1.2054684}
  {\path{doi:10.1149/1.2054684}}.

\bibitem{zulke2021parametrisation}
A.~Z{\"u}lke, I.~Korotkin, J.~M. Foster, M.~Nagarathinam, H.~Hoster,
  G.~Richardson, Parametrisation and use of a predictive dfn model for a
  high-energy nca/gr-siox battery, Journal of The Electrochemical Society
  168~(12) (2021) 120522.

\bibitem{schmitt2023full}
C.~Schmitt, M.~Gerle, D.~Kopljar, K.~A. Friedrich, Full parameterization study
  of a high-energy and high-power li-ion cell for physicochemical models,
  Journal of The Electrochemical Society 170~(7) (2023) 070509.

\bibitem{ecker2015parameterization}
M.~Ecker, T.~K.~D. Tran, P.~Dechent, S.~K{\"a}bitz, A.~Warnecke, D.~U. Sauer,
  Parameterization of a physico-chemical model of a lithium-ion battery: I.
  determination of parameters, Journal of The Electrochemical Society 162~(9)
  (2015) A1836.

\bibitem{ecker2015parameterizationb}
M.~Ecker, S.~K{\"a}bitz, I.~Laresgoiti, D.~U. Sauer, Parameterization of a
  physico-chemical model of a lithium-ion battery: Ii. model validation,
  Journal of The Electrochemical Society 162~(9) (2015) A1849.

\bibitem{BrosaPlanella2022}
F.~B. Planella, W.~Ai, A.~M. Boyce, A.~Ghosh, I.~Korotkin, S.~Sahu, V.~Sulzer,
  R.~Timms, T.~G. Tranter, M.~Zyskin, S.~J. Cooper, J.~S. Edge, J.~M. Foster,
  M.~Marinescu, B.~Wu, G.~Richardson,
  \href{https://dx.doi.org/10.1088/2516-1083/ac7d31}{A continuum of
  physics-based lithium-ion battery models reviewed}, Progress in Energy 4~(4)
  (2022) 042003.
\newblock \href {https://doi.org/10.1088/2516-1083/ac7d31}
  {\path{doi:10.1088/2516-1083/ac7d31}}.
\newline\urlprefix\url{https://dx.doi.org/10.1088/2516-1083/ac7d31}

\bibitem{moura2016battery}
S.~J. Moura, F.~B. Argomedo, R.~Klein, A.~Mirtabatabaei, M.~Krstic, Battery
  state estimation for a single particle model with electrolyte dynamics, IEEE
  Transactions on Control Systems Technology 25~(2) (2016) 453--468.

\bibitem{guo2010single}
M.~Guo, G.~Sikha, R.~E. White, Single-particle model for a lithium-ion cell:
  Thermal behavior, Journal of The Electrochemical Society 158~(2) (2010) A122.

\bibitem{marquis2019asymptotic}
S.~G. Marquis, V.~Sulzer, R.~Timms, C.~P. Please, S.~J. Chapman, An asymptotic
  derivation of a single particle model with electrolyte, Journal of The
  Electrochemical Society 166~(15) (2019) A3693.

\bibitem{richardson2020generalised}
G.~Richardson, I.~Korotkin, R.~Ranom, M.~Castle, J.~Foster, Generalised single
  particle models for high-rate operation of graded lithium-ion electrodes:
  Systematic derivation and validation, Electrochimica Acta 339 (2020) 135862.

\bibitem{Wang_2022}
A.~A. Wang, S.~E.~J. O’Kane, F.~B. Planella, J.~L. Houx, K.~O’Regan,
  M.~Zyskin, J.~Edge, C.~W. Monroe, S.~J. Cooper, D.~A. Howey, E.~Kendrick,
  J.~M. Foster, \href{https://dx.doi.org/10.1088/2516-1083/ac692c}{Review of
  parameterisation and a novel database (liiondb) for continuum li-ion battery
  models}, Progress in Energy 4~(3) (2022) 032004.
\newblock \href {https://doi.org/10.1088/2516-1083/ac692c}
  {\path{doi:10.1088/2516-1083/ac692c}}.
\newline\urlprefix\url{https://dx.doi.org/10.1088/2516-1083/ac692c}

\bibitem{em1}
R.~Tian, P.~J. King, J.~Coelho, S.-H. Park, D.~V. Horvath, V.~Nicolosi,
  C.~O'Dwyer, J.~N. Coleman, Using chronoamperometry to rapidly measure and
  quantitatively analyse rate-performance in battery electrodes, Journal of
  Power Sources 468 (2020) 228220.

\bibitem{Bard_2001_book}
A.~J. Bard, L.~R. Faulkner, Electrochemical Methods: Fundamentals and
  Applications, 2nd Edition, Wiley, 2001.

\bibitem{bark}
J.~Barker, R.~Pynenburg, R.~Koksbang, M.~Saidi, An electrochemical
  investigation into the lithium insertion properties of lixcoo2,
  Electrochimica acta 41~(15) (1996) 2481--2488.

\bibitem{Weppner1977}
W.~Weppner, R.~A. Huggins, Determination of the kinetic parameters of
  mixed-conducting electrodes and application to the system li$_3$6sb, J.
  Electrochem. Soc 124 (1977) 43.

\bibitem{PMID:37085556}
Y.-C. Chien, H.~Liu, A.~S. Menon, W.~R. Brant, D.~Brandell, M.~J. Lacey,
  \href{https://europepmc.org/articles/PMC10121696}{Rapid determination of
  solid-state diffusion coefficients in li-based batteries via intermittent
  current interruption method}, Nature communications 14~(1) (2023) 2289.
\newblock \href {https://doi.org/10.1038/s41467-023-37989-6}
  {\path{doi:10.1038/s41467-023-37989-6}}.
\newline\urlprefix\url{https://europepmc.org/articles/PMC10121696}

\bibitem{yin}
L.~Yin, Z.~Geng, Y.-C. Chien, T.~Thiringer, M.~J. Lacey, A.~M. Andersson,
  D.~Brandell, Implementing intermittent current interruption into li-ion cell
  modelling for improved battery diagnostics, Electrochimica Acta 427 (2022)
  140888.

\bibitem{kim2020}
T.~Kim, W.~Choi, H.-C. Shin, J.-Y. Choi, J.~M. Kim, M.-S. Park, W.-S. Yoon,
  Applications of voltammetry in lithium ion battery research, Journal of
  Electrochemical Science and Technology 11~(1) (2020) 14--25.

\bibitem{yu1999}
P.~Yu, B.~N. Popov, J.~A. Ritter, R.~E. White, Determination of the lithium ion
  diffusion coefficient in graphite, Journal of The Electrochemical Society
  146~(1) (1999) 8.

\bibitem{lee2022}
H.~Lee, S.~Yang, S.~Kim, J.~Song, J.~Park, C.-H. Doh, Y.-C. Ha, T.-S. Kwon,
  Y.~M. Lee, Understanding the effects of diffusion coefficient and exchange
  current density on the electrochemical model of lithium-ion batteries,
  Current Opinion in Electrochemistry 34 (2022) 100986.

\bibitem{chen2022}
Y.~Chen, J.~Key, K.~O'regan, T.~Song, Y.~Han, E.~Kendrick, Revealing the
  rate-limiting electrode of lithium batteries at high rates and mass loadings,
  Chemical Engineering Journal 450 (2022) 138275.

\bibitem{doi:10.1080/14786440109462590}
H.~J. Sand, \href{https://doi.org/10.1080/14786440109462590}{Iii. on the
  concentration at the electrodes in a solution, with special reference to the
  liberation of hydrogen by electrolysis of a mixture of copper sulphate and
  sulphuric acid}, The London, Edinburgh, and Dublin Philosophical Magazine and
  Journal of Science 1~(1) (1901) 45--79.
\newblock \href {https://doi.org/10.1080/14786440109462590}
  {\path{doi:10.1080/14786440109462590}}.
\newline\urlprefix\url{https://doi.org/10.1080/14786440109462590}

\bibitem{Nickol_2020}
A.~Nickol, T.~Schied, C.~Heubner, M.~Schneider, A.~Michaelis, M.~Bobeth,
  G.~Cuniberti, \href{https://dx.doi.org/10.1149/1945-7111/ab9404}{Gitt
  analysis of lithium insertion cathodes for determining the lithium diffusion
  coefficient at low temperature: Challenges and pitfalls}, Journal of The
  Electrochemical Society 167~(9) (2020) 090546.
\newblock \href {https://doi.org/10.1149/1945-7111/ab9404}
  {\path{doi:10.1149/1945-7111/ab9404}}.
\newline\urlprefix\url{https://dx.doi.org/10.1149/1945-7111/ab9404}

\bibitem{horner2021}
J.~S. Horner, G.~Whang, D.~S. Ashby, I.~V. Kolesnichenko, T.~N. Lambert, B.~S.
  Dunn, A.~A. Talin, S.~A. Roberts,
  \href{https://doi.org/10.1021/acsaem.1c02218}{Electrochemical modeling of
  gitt measurements for improved solid-state diffusion coefficient evaluation},
  ACS Applied Energy Materials 4~(10) (2021) 11460--11469.
\newblock \href {http://arxiv.org/abs/https://doi.org/10.1021/acsaem.1c02218}
  {\path{arXiv:https://doi.org/10.1021/acsaem.1c02218}}, \href
  {https://doi.org/10.1021/acsaem.1c02218} {\path{doi:10.1021/acsaem.1c02218}}.
\newline\urlprefix\url{https://doi.org/10.1021/acsaem.1c02218}

\bibitem{IVANISHCHEV2017479}
A.~V. Ivanishchev, A.~V. Ushakov, I.~A. Ivanishcheva, A.~V. Churikov, A.~V.
  Mironov, S.~S. Fedotov, N.~R. Khasanova, E.~V. Antipov,
  \href{https://www.sciencedirect.com/science/article/pii/S0013468617302591}{Structural
  and electrochemical study of fast li diffusion in li3v2(po4)3-based electrode
  material}, Electrochimica Acta 230 (2017) 479--491.
\newblock \href
  {https://doi.org/https://doi.org/10.1016/j.electacta.2017.02.009}
  {\path{doi:https://doi.org/10.1016/j.electacta.2017.02.009}}.
\newline\urlprefix\url{https://www.sciencedirect.com/science/article/pii/S0013468617302591}

\bibitem{Zeng2013}
Y.~Zeng, P.~Albertus, R.~Klein, N.~Chaturvedi, A.~Kojic, M.~Z. Bazant,
  J.~Christensen, \href{https://dx.doi.org/10.1149/2.102309jes}{Efficient
  conservative numerical schemes for 1d nonlinear spherical diffusion equations
  with applications in battery modeling}, Journal of The Electrochemical
  Society 160~(9) (2013) A1565.
\newblock \href {https://doi.org/10.1149/2.102309jes}
  {\path{doi:10.1149/2.102309jes}}.
\newline\urlprefix\url{https://dx.doi.org/10.1149/2.102309jes}

\bibitem{Chen_2020}
C.-H. Chen, F.~B. Planella, K.~O’Regan, D.~Gastol, W.~D. Widanage,
  E.~Kendrick, \href{https://dx.doi.org/10.1149/1945-7111/ab9050}{Development
  of experimental techniques for parameterization of multi-scale lithium-ion
  battery models}, Journal of The Electrochemical Society 167~(8) (2020)
  080534.
\newblock \href {https://doi.org/10.1149/1945-7111/ab9050}
  {\path{doi:10.1149/1945-7111/ab9050}}.
\newline\urlprefix\url{https://dx.doi.org/10.1149/1945-7111/ab9050}

\bibitem{Kendrick_C20_data}
{The E. Kendrick group, Birmingham University, unpublished data}.

\bibitem{Korotkin_2021}
I.~Korotkin, S.~Sahu, S.~E.~J. O'Kane, G.~Richardson, J.~M. Foster,
  \href{https://doi.org/10.1149/1945-7111/ac085f}{{DandeLiion} v1: An extremely
  fast solver for the newman model of lithium-ion battery (dis)charge}, Journal
  of The Electrochemical Society 168~(6) (2021) 060544.
\newblock \href {https://doi.org/10.1149/1945-7111/ac085f}
  {\path{doi:10.1149/1945-7111/ac085f}}.
\newline\urlprefix\url{https://doi.org/10.1149/1945-7111/ac085f}

\bibitem{Carslaw1959-to}
H.~S. Carslaw, J.~C. Jaeger, Conduction of heat in solids, 2nd Edition, Oxford
  University Press, London, England, 1959.

\bibitem{SUBRAMANIAN2001385}
V.~R. Subramanian, R.~E. White,
  \href{https://www.sciencedirect.com/science/article/pii/S037877530000656X}{New
  separation of variables method for composite electrodes with galvanostatic
  boundary conditions}, Journal of Power Sources 96~(2) (2001) 385--395.
\newblock \href {https://doi.org/https://doi.org/10.1016/S0378-7753(00)00656-X}
  {\path{doi:https://doi.org/10.1016/S0378-7753(00)00656-X}}.
\newline\urlprefix\url{https://www.sciencedirect.com/science/article/pii/S037877530000656X}

\bibitem{Liron71}
N.~Liron, \href{https://doi.org/10.1137/0502010}{Some infinite sums}, SIAM
  Journal on Mathematical Analysis 2~(1) (1971) 105--112.
\newblock \href {http://arxiv.org/abs/https://doi.org/10.1137/0502010}
  {\path{arXiv:https://doi.org/10.1137/0502010}}, \href
  {https://doi.org/10.1137/0502010} {\path{doi:10.1137/0502010}}.
\newline\urlprefix\url{https://doi.org/10.1137/0502010}

\bibitem{cacuci2003sensitivity}
D.~G. Cacuci, \href{https://doi.org/10.1201/9780203498798}{Sensitivity \&
  Uncertainty Analysis, Volume 1: Theory}, 1st Edition, Chapman and Hall/CRC,
  2003.
\newblock \href {https://doi.org/10.1201/9780203498798}
  {\path{doi:10.1201/9780203498798}}.
\newline\urlprefix\url{https://doi.org/10.1201/9780203498798}

\bibitem{KUZHIYIL2025125221}
J.~A. Kuzhiyil, T.~Damoulas, F.~B. Planella, W.~D. Widanage,
  \href{https://www.sciencedirect.com/science/article/pii/S0306261924026059}{Lithium-ion
  battery degradation modelling using universal differential equations:
  Development of a cost-effective parameterisation methodology}, Applied Energy
  382 (2025) 125221.
\newblock \href
  {https://doi.org/https://doi.org/10.1016/j.apenergy.2024.125221}
  {\path{doi:https://doi.org/10.1016/j.apenergy.2024.125221}}.
\newline\urlprefix\url{https://www.sciencedirect.com/science/article/pii/S0306261924026059}

\end{thebibliography}

\end{document}